\documentclass[twocolumn]{aastex61}
\usepackage{graphicx}	
\usepackage{amsmath}	
\usepackage{amssymb}	
\usepackage{enumitem}

\defcitealias{lehner13}{L13}
\defcitealias{wotta16}{W16}
\defcitealias{wotta18}{W18}
\defcitealias{fumagalli16}{FOP16}

\newcommand{\ssz}{224}
\newcommand{\tsz}{263}
\newcommand{\uvesobs}{12}
\newcommand{\gbmgii}{113}

\def\nodata{ ~$\cdots$~ }

\newcommand{\mzabs}{z_{\rm abs}}

\newcommand{\cmm}{cm$^{-2}$}

\newcommand{\lya}{Ly$\alpha$}

\newcommand{\lyb}{Ly$\beta$}
\newcommand{\lyg}{Ly$\gamma$}
\newcommand{\nhi}{$N_{\rm H\,I}$}

\newcommand{\noi}{$N_{\rm O\,I}$}
\newcommand{\nmgii}{$N_{\rm Mg\,II}$}
\newcommand{\ncii}{$N_{\rm C\,II}$}
\newcommand{\nciii}{$N_{\rm C\,III}$}

\newcommand{\mnhi}{N_{\rm H\,I}}

\newcommand{\lnhi}{$\log N_{\rm H\,I}$}

\newcommand{\mlnmgii}{\log N_{\rm Mg\,II}}

\newcommand{\mlnhi}{\log N_{\rm H\,I}}

\newcommand{\xh}{{\rm [X/H]}}
\newcommand{\km}{${\rm km\,s}^{-1}$}

\newcommand{\hst}{{\em HST}}
\newcommand{\fuse}{{\em FUSE}}

\newcommand{\mdv}{\Delta v_{90}}
\newcommand{\dv}{$\Delta v_{90}$}

\newcommand{\hi}{H$\;${\small\rm I}\relax}

\newcommand{\nni}{N$\;${\small\rm I}\relax}

\newcommand{\cii}{C$\;${\small\rm II}\relax}
\newcommand{\ciii}{C$\;${\small\rm III}\relax}
\newcommand{\civ}{C$\;${\small\rm IV}\relax}
\newcommand{\nii}{N$\;${\small\rm II}\relax}

\newcommand{\nv}{N$\;${\small\rm V}\relax}
\newcommand{\oi}{O$\;${\small\rm I}\relax}
\newcommand{\oii}{O$\;${\small\rm II}\relax}
\newcommand{\oiii}{O$\;${\small\rm III}\relax}

\newcommand{\ovi}{O$\;${\small\rm VI}\relax}

\newcommand{\sii}{S$\;${\small\rm II}\relax}
\newcommand{\siii}{Si$\;${\small\rm II}\relax}
\newcommand{\siiii}{Si$\;${\small\rm III}\relax}

\newcommand{\siiv}{Si$\;${\small\rm IV}\relax}
\newcommand{\siv}{S$\;${\small\rm IV}\relax}
\newcommand{\mgii}{Mg$\;${\small\rm II}\relax}

\newcommand{\sv}{S$\;${\small\rm V}\relax}
\newcommand{\svi}{S$\;${\small\rm VI}\relax}
\newcommand{\feii}{Fe$\;${\small\rm II}\relax}

\newcommand{\hit}{H$\;${\scriptsize\rm I}\relax}

\newcommand{\neviiit}{Ne$\;${\scriptsize\rm VIII}\relax}
\newcommand{\mgxt}{Mg$\;${\scriptsize\rm X}\relax}

\newcommand{\ciit}{C$\;${\scriptsize\rm II}\relax}

\newcommand{\ciiit}{C$\;${\scriptsize\rm III}\relax}
\newcommand{\civt}{C$\;${\scriptsize\rm IV}\relax}
\newcommand{\nnit}{N$\;${\scriptsize\rm I}\relax}
\newcommand{\niit}{N$\;${\scriptsize\rm II}\relax}
\newcommand{\niiit}{N$\;${\scriptsize\rm III}\relax}
\newcommand{\nivt}{N$\;${\scriptsize\rm IV}\relax}
\newcommand{\nvt}{N$\;${\scriptsize\rm V}\relax}
\newcommand{\oit}{O$\;${\scriptsize\rm I}\relax}
\newcommand{\oiit}{O$\;${\scriptsize\rm II}\relax}
\newcommand{\oiiit}{O$\;${\scriptsize\rm III}\relax}
\newcommand{\oivt}{O$\;${\scriptsize\rm IV}\relax}
\newcommand{\ovt}{O$\;${\scriptsize\rm V}\relax}
\newcommand{\ovit}{O$\;${\scriptsize\rm VI}\relax}

\newcommand{\mgiit}{Mg$\;${\scriptsize\rm II}\relax}
\newcommand{\siit}{S$\;${\scriptsize\rm II}\relax}
\newcommand{\siiit}{Si$\;${\scriptsize\rm II}\relax}
\newcommand{\svt}{S$\;${\scriptsize\rm V}\relax}
\newcommand{\siiiit}{Si$\;${\scriptsize\rm III}\relax}

\newcommand{\Siiit}{S$\;${\scriptsize\rm III}\relax}
\newcommand{\siivt}{Si$\;${\scriptsize\rm IV}\relax}
\newcommand{\sivt}{S$\;${\scriptsize\rm IV}\relax}
\newcommand{\svit}{S$\;${\scriptsize\rm VI}\relax}
\newcommand{\feiit}{Fe$\;${\scriptsize\rm II}\relax}

\newcommand{\feiiit}{Fe$\;${\scriptsize\rm III}\relax}

\shortauthors{Lehner et al.}
\shorttitle{The COS CGM Compendium (CCC).  I: Survey Design and Initial Results}

\begin{document}

\title{The COS CGM Compendium (CCC). I: Survey Design and Initial Results}

\author{Nicolas Lehner}
\affiliation{Department of Physics, University of Notre Dame, Notre Dame, IN 46556, USA}
\author{Christopher B. Wotta}
\affiliation{Department of Physics, University of Notre Dame, Notre Dame, IN 46556, USA}
\author{J. Christopher Howk}
\affiliation{Department of Physics, University of Notre Dame, Notre Dame, IN 46556, USA}
\author{John M. O'Meara}
\affiliation{Department of Chemistry and Physics, Saint Michael's College, Colchester, VT 05439, USA}
\author{Benjamin D. Oppenheimer}
\affiliation{CASA, Department of Astrophysical and Planetary Sciences, University of Colorado, Boulder, CO 80309, USA}
\author{Kathy L.~Cooksey}
\affiliation{Department of Physics and Astronomy, University of Hawai`i at Hilo, HI 96720, USA}

\begin{abstract}
We present a neutral hydrogen-selected absorption-line survey of gas with \hi\ column densities $15 < \mlnhi < 19$  at $z\la 1$ using the Cosmic Origins Spectrograph on the {\em Hubble Space Telescope}. Our main aim is to determine the metallicity distribution of these absorbers. Our sample consists of \ssz\ absorbers selected on the basis of their \hi\ absorption strength. Here we discuss the properties of our survey and the immediate empirical results. We find singly and doubly ionized metal species and \hi\ typically have similar velocity profiles, implying they probe gas in the same or similar environments. The ionic ratios (e.g., \ncii/\nciii, \noi/\ncii) indicate the gas in these absorbers is largely ionized, and the ionization conditions are quite comparable across the sampled \nhi\ range. The Doppler parameters of the \hi\ imply $T \la 5\times 10^4$ K on average, consistent with the gas being photoionized. The \mgii\ column densities span $>2$ orders of magnitude at any given \nhi, indicating a wide range of metallicities (from solar to $<1/100$ solar). In the range  $16.2\la \mlnhi \la 17$, there is a gap in the \nmgii\ distribution corresponding to gas with $\sim 10\%$ solar metallicity, consistent with the gap seen in the previously identified bimodal metallicity distribution in this column density regime.  Less than $3\%$ of the absorbers in our sample show no detectable metal absorption, implying truly-pristine gas at $z\la 1$ is uncommon. We find $\langle [$\feii/\mgii$]\rangle = -0.4 \pm 0.3$, and since $\alpha$-enhancement can affect this ratio, dust depletion is extremely mild. 
\end{abstract}

\keywords{quasars: absorption lines  ---   galaxies: halos --- abundances}

\section{Introduction}\label{s-intro}
Modern theory and cosmological simulations agree that the star formation of galaxies and the properties of their circumgalactic medium (CGM) should be intimately connected \citep[see recent review by][]{tumlinson17}. This is especially true 
for the flows through the CGM: feedback from star formation is understood to drive outflows that carry mass and metals away from galaxies, while infall from the intergalactic medium (IGM) is thought to bring in fresh gas to fuel on-going star formation. Without significant feedback, most baryons would cool into the centers of halos to form prodigious quantities of stars \citep[e.g.,][]{white78,keres09}, but with feedback, the baryon content of stars and cold gas in galaxies can be matched by driving matter into the CGM and beyond \citep[e.g.,][]{fukugita98,conroy09}. Similarly, without continued infall of IGM material, star-forming galaxies would consume their interstellar gas in $\sim$1 Gyr (e.g., \citealt{rocha-pinto00}). Gas accretion may also control, in part, the evolution of the elemental abundances in galaxies and could also play a role in the mass-metallicity relationship \citep[e.g.,][]{kacprzak16}. These exchanges of matter through the CGM thus play critical roles in the evolution of galaxies. In our Cosmic Origins Spectrograph (COS) CGM Compendium (CCC) presented in this paper, we aim to directly explore how a specific property of the CGM --- the metallicity --- is distributed throughout cosmic time and environments.

Observationally, outflows have been characterized at various redshifts and appear ubiquitous in the universe \citep[e.g.,][]{shapley03,steidel10,weiner09,martin13,rubin14}. However, direct observational evidence for cold  gas accretion  (colder than the virial temperature, i.e. gas at $ \sim 10^{4}$--$10^{5}$ K) has been more difficult to discover. There is some evidence for gas accretion onto the Milky Way and other nearby galaxies (e.g., \citealt{wakker01,fraternali08,fox16,fox18}). Recent integral-field spectroscopic observations and studies of redshift-space distortions also suggest some evidence of inflows at low and high redshifts  \citep[e.g.,][]{fumagalli16,bielby17,cantalupo17,turner17}. The $z=0$ observations indicate that gas can be accreted via other means than cold flow accretion, while the higher redshift results provide some support to cold flow accretion as observed in simulations, although none of these observational results show  yet conclusively and unambiguously the existence of IGM gaseous filaments feeding galaxies. Another indirect way to probe cold flow accretion (and more generally gas flows in and out of galaxies) is via QSO absorbers that probe the CGM of galaxies. According to cosmological simulations \citep[e.g.,][]{fumagalli11a,vandevoort12a,vandevoort12b,faucher-giguere15,hafen17}, the covering fraction of the cold streams in these simulations appears to peak in a \hi\ column density (\nhi) regime known as the (partial) Lyman Limit systems (pLLSs and LLSs). Using QSO spectra, we can search specifically for these absorbers and empirically characterize their properties. 

In this and subsequent papers, we adhere to the following definition for the various absorbers studied in our survey. The pLLSs and LLSs have \hi\ column densities  $16.2 \la \mlnhi <17.2 $ and $17.2 \le \mlnhi <19$, respectively. The $\mlnhi = 17.2$ cutoff corresponds to an optical depth at the Lyman limit $\tau_{\rm LL} = 1$ ($\tau_{\rm LL} \equiv \mnhi\ \sigma_{\rm H\,I}$, where $\sigma_{\rm H\,I} = 6.30 \times 10^{-18}$ cm$^2$ is the absorption cross section of a hydrogen atom at the Lyman limit, see \citealt{spitzer78}). The 16.2 dex lower bound of the pLLSs is more arbitrary, but corresponds to $\tau_{\rm LL} = 0.1$, which creates a break at the Lyman limit that is still visible in moderate and high signal-to-noise ratio (SNR) spectra (see \citealt{lehner13}, hereafter \citetalias{lehner13}). This definition of LLSs differs from our previous surveys but follows more closely the generally adopted definition of LLSs and is motivated by the finding of \citet{wotta16} (hereafter \citetalias{wotta16}), which shows a tentative difference in the metallicity distribution between the pLLSs and LLSs at $z\la 1$ (see below). We also use the standard definition for the damped \lya\ absorbers (DLAs) that have $\mlnhi \ge 20.3$ and super-LLSs (SLLSs, a.k.a. sub-DLAs) with \nhi\ in the range $19.0 \le \mlnhi <20.3$. The absorbers with $15 \la \mlnhi <16.2$ have no proper definition and we simply define these absorbers as the strong \lya\ forest systems (SLFSs). We, however, emphasize the SLFSs are more related to the diffuse CGM than the IGM at $z<1$. Indeed, several studies of the  galaxy--absorber two-point cross-correlation function have shown significant clustering between galaxies and SLFSs while a weak or absent clustering signal for weaker \nhi\ absorbers \citep[e.g.,][]{lanzetta95,penton02,bowen02,chen05,prochaska11a,prochaska11c,tejos14}, pointing to a strong physical connection between SLFSs (and other stronger  \hi\ absorbers) and galaxies. 

While the diffuse gas probed by SLFSs, pLLSs, LLSs cannot be yet directly imaged, the properties of these absorbers can inform us on their origins and hence help us characterize the gas in the CGM of galaxies at low redshift. One of the key properties is the metallicity since the enrichment levels of the gas help in differentiating between the plausible origins of the gas (\citealt{ribaudo11,fumagalli11a}; \citetalias{lehner13}). For example, we can differentiate  accretion from the IGM (that is metal-poor) from the more  metal-rich gas produced  by galaxies (which we emphasize can be outflowing or inflowing). Determining the metallicity distribution of the CGM gas also informs us about the fraction that has remained mostly untouched by the successive episodes of star formation in galaxies over billions of years. 

While metallicity estimates nearly always require large ionization corrections in the $15 < \mlnhi < 19$ range, several independent studies have shown that such corrections can be well constrained by the broad range of ions despite an uncertain EUV ionizing background (\citealt{howk09}; \citetalias{lehner13}; \citealt{crighton13a,fumagalli16}; \citetalias{wotta16}). The uncertainty in the ionizing background can lead to an absolute uncertainty in the metallicity of a factor 2--3 (\citealt{howk09}; \citetalias{wotta16}; \citealt{fumagalli16}). Though it is not precision cosmology, it is accurate enough to separate low from high metallicities as well as enable a determination of the metallicity probability distribution function of these absorbers. We bear in mind that this error level on the metallicity of the absorbers is comparable to the magnitude of uncertainty on chemical abundances determined from emission lines in galaxy spectra (e.g., \citealt{berg16}). However, we also note  that the uncertainties in the metallicities for a given ionizing background are also typically quite small (depending on the exact constraints provided by the available ions; see \citetalias{lehner13}; \citealt{fumagalli16}).  

Over the last few years, our team has led several surveys of the metallicities of the pLLSs and LLSs at $z\la 1$ (\citealt{lehner09}; \citetalias{lehner13}; \citealt{ribaudo11,tripp11}; \citetalias{wotta16}) and $z\sim 2$--4 (\citealt{lehner16}, and see also \citealt{fumagalli16,cooper15,glidden16}). One of our main findings has been the first evidence of widespread low metallicity gas in the pLLS and LLS regime that is not observed in higher column density absorbers at similar redshifts. At $z\la 1$ there is indeed a clear prevalence of very low metallicity pLLSs with $\xh \equiv \log N_{\rm X}/N_{\rm H} - \log {\rm X/H}_\sun \la -1.4$ (where X is an $\alpha$-element such as Mg, O, Si) with $43\% \pm 8\%$ of the pLLSs with these low metallicities, while $<10\%$ of the DLAs and SLLSs have $\xh < -1.4$ (where X is now a $\alpha$-element, Zn, or Fe corrected for depletion or $\alpha/{\rm Fe}$ enhancement) \citepalias{wotta16}. The pLLSs and LLSs therefore uniquely probe metal-poor gas in relatively dense (denser than the IGM probed by weak \lya\ forest absorbers)  environments in the  universe at low and high redshift (\citetalias{lehner13,wotta16}; \citealt{lehner16,fumagalli16,cooper15,glidden16}). Another major finding from these initial surveys was that the shape of the metallicity distribution  of the pLLSs at $z\la 1$ appears to be bimodal with an equal number of pLLSs below and above $\xh = -1$, a functional form only observed at $z<1$ (at $z>2$, the metallicity distribution appears to be unimodal for similar absorbers, see \citealt{lehner16}). 

These empirical results provide the first strong evidence of widespread very low metallicity gas around $z\la 1$ galaxies that may possibly accrete onto galaxies, although we emphasize that there is not yet any direct evidence that this gas is actually accreting. Our initial survey consisted of 28 absorbers (23 pLLSs and 5 LLSs) \citepalias{lehner13}. Our second survey doubled the size of our initial sample (44 pLLSs and 11 LLSs), confirming our initial results, and tentatively demonstrating a possible change in the shape of the metallicity distribution between the pLLSs and LLSs and a lower frequency of metal poor LLSs compared to pLLSs \citepalias{wotta16}.  Our surveys have increased by an order of magnitude the number of pLLSs and LLSs where the metallicities have been estimated at $z\la 1$ compared to the status prior to the installation of the COS onboard the {\it Hubble Space Telescope} (\hst) (see compilation by \citealt{lehner09} and references therein). However, the sample of absorbers is still relatively small, especially when we treat the pLLSs and LLSs separately. Furthermore to  effectively probe the level of the IGM/CGM enrichment requires that we not only increase the number of strong \hi\ absorbers, but that we also target lower \nhi\ absorbers. 

We were awarded a  {\it Hubble Space Telescope} (\hst) Legacy program to produce the largest survey to date of \hi-selected absorbers at $z\la 1$ using the spectra observed with the high-resolution mode (G130M and G160M gratings) of COS available at the \textit{Barbara A. Mikulski Archive for Space Telescopes} (MAST). This archive is the richest database of UV QSO spectra with sufficient quality (spectral resolution and SNR) that can permit accurate estimations of the column densities of metal and \hi\ transitions. As we detail in this paper, \nhi\ can be well determined by modeling the entire Lyman series (which is accessible for $z_{\rm em}\ga 0.2$ QSOs in the COS bandpass). The restframe UV, FUV, and EUV wavelengths covered by the COS spectra include a large number of metal lines with a wide range of strengths and ionization states, and additional NUV transitions (\mgii\ $\lambda \lambda$2796, 2803, \feii\ $\lambda \lambda$2382, 2600) can be observed from ground-based telescopes for the redshifts probed by our survey.  

At $z\la 1$, we note that any absorbers with $\mnhi \ga 10^{15.2}$ \cmm\ are in a relative ``sweet spot'' for an unbiased metallicity study because both the metal and \hi\ column densities can be accurately estimated to provide a reliable census of the metallicity distribution. The SNRs of the COS spectra are, however, not high enough to pursue an unbiased survey of the metallicity in the more diffuse gas probed by \lya\ forest absorbers with $\mlnhi < 15$ at $z\la 1$. 

Here we present our CCC survey of \hi-selected absorbers with column densities in the range $15<\mlnhi < 19$ at $z\la 1$ aimed to determine the metallicities of these absorbers. We used the \hst\ COS G130M and G160M archive that we supplemented with additional spectra of \mgii\ and \feii\ obtained from the  High Resolution Echelle Spectrometer (HIRES) on Keck I and the Ultraviolet and Visual Echelle Spectrograph (UVES) on the Very Large Telescope (VLT). Our COS G130M and G160M survey presented here comprises \ssz\ absorbers, and our total sample for the metallicity study has \tsz\ absorbers (the additional absorbers that were primarily observed with other instruments or gratings). In this paper, we present the design survey, observations, and data reductions (\S\ref{s-present}), the search for strong \hi\ absorption in the COS spectra (\S\ref{s-search}) and column density and kinematics measurements of the \ssz\ absorbers (\S\ref{s-estimate-col}), and the immediate implications from these measurements (\S\ref{s-results}).  We summarize our main results in \S\ref{s-summary}. In the subsequent papers, we will present the grids of ionization models combined with a Bayesian formalism and Markov Chain Monte Carlo (MCMC) techniques to robustly determine the metallicities (and errors) of the absorbers (paper II), the metallicity probability distribution of the pLLSs and LLSs (paper II) and SLFSs (paper III), and the properties of the high ions in these absorbers (paper IV). 

\section{Survey Design, Database, and Observations}\label{s-present}
\subsection{Sample Selection Criteria}\label{s-criteria}

Our main goal for this survey is to estimate the metallicities of absorbers in the \hst\ COS G130M and G160M archive. This requires that we select absorbers for which it is possible to derive the column densities of \hi\ and at least some metal ions (preferably  $\alpha$ elements like O, Mg, Si). In order not to bias our survey toward low or high metallicities, this requires that 1) we identify absorbers based on their \hi\ content alone,  and 2) we must also be sensitive to metal-line absorption. Since the typical SNRs of the COS G130M and G160M data are $\la 30$ (and often $\la 10$) per resolution element, we have determined --- using guidance from Cloudy simulations --- that this requires $\mlnhi \ga 15.2$ at $z\la 1$ to be sensitive to metallicities $\xh \la -1$ (see also below).\footnote{As we detail below, we can also use the \mgiit\ $\lambda \lambda$2796, 2803 doublet obtained from ground-based telescopes to derive the metallicity, but for $\mlnhi \ga 15.2$, it becomes prohibitively expensive since it requires SNR\,$\ga 100$ in the high-resolution mode of a 8--10 m telescope.} Therefore our survey is \hi-selected so that $\mlnhi \ga 15.2$. In \S\ref{s-search}, we provide more details on the search for the absorbers. 

We note that a more subtle bias could, in principle, be introduced in our survey if a large number of absorbers were initially targeted because they were known {\it a priori} to be metal rich or poor systems. To the best of our knowledge and after scanning through the description of the various \hst\ programs that originally obtained the data, we feel that our survey is not adversely affected in this way; i.e., most of the absorbers were serendipitously observed in the spectra that targeted other science goals (e.g., study of the Milky Way halo, Local group gas, the diffuse IGM). The only exception that we are aware of is the metal-poor absorber at $z=0.660356$ toward J131956.23+272808.2 (a.k.a, CSO-0873 or TON153), which was selected based on its \mgii\ absorption \citep{kacprzak12,churchill12}, as already noted in \citetalias{lehner13}. Here we provide a new estimate of its \hi\ column density, but do not include this absorber in our survey. Several absorbers were also found in the spectra from the COS-Halos survey \citep{werk14,werk13,tumlinson13,tumlinson11a}, but the vast majority of those are not the absorbers that were initially targeted to probe  the CGM of their galaxy sample. In fact we have only 9 absorbers in common with the targeted COS-Halos absorbers. This number is not larger because 1) several of the COS-Halos absorbers have either lower or high \nhi\ than probed by our survey; 2) several of the COS-Halos absorbers have $\mzabs < 0.2$, which is outside our redshift search range (see \S\ref{s-search}); 3) we did not use the subsequent COS G140L spectra obtained by the COS-Halos team to determine \nhi\ for absorbers that had only a lower limit on \nhi\ from the G130M and G160M spectra; and 4) for a couple of absorbers we could not derive a reliable \nhi. We will provide a detailed comparison of our results with those from COS-Halos in paper III.

We note that our identification and analysis of absorbers, including estimations of their metallicities, is somewhat different than that adopted by COS-Halos and by several other surveys of high column density absorbers. Where possible, we separate individual absorbers on the basis of our ability to cleanly resolve them from one another in the \hst /COS spectroscopy. We do not impose a minimum velocity or redshift window across which we assume all absorption is associated with a single system. This is related to the discussion above, in that we only associate metal line absorption with its directly associated \hi\ absorption component.  We note that this choice amounts to a different working definition of what constitutes an absorption system. As we discuss in \S\ref{s-prox}, absorbers closely separated in redshift space ($\Delta v_c \le 500$ \km) consist only of about 13\% of the sample analyzed in this paper. 

In our search, we also did not reject {\it a priori} proximate absorbers, i.e., absorbers spatially close in redshift space to the background QSO (in this work defined as $\Delta v < 3000$ \km). The proximate absorbers constitute a small fraction of the sample with only 5\% of the absorbers being proximate (see \S\ref{s-prox}). In papers II and III, we will note the influence of these proximate absorbers on our results as needed, but they represent a small portion of the sample and we will demonstrate they have little impact on the overall properties we derive from our survey.  

\subsection{COS Data and Database}\label{s-data}

Our survey makes use of the extended archival UV spectroscopy of AGN and QSOs taken with the highest-resolution modes of the COS instrument. General information on the design and performance of COS can be found in \citet{green12}. We used only data taken with the COS G130M and G160M gratings, with spectral resolutions $R\approx 17,000$ (or about 15--20 \km). All the COS G130M and G160M data were retrieved from MAST. Our work started prior to the release of the HST Spectroscopic Legacy Archive (HSLA, \citealt{peeples17}) reduction of the COS archival data. Thus we searched the MAST archive for observations of AGN/QSOs taken with the G130M and/or G160M observations using a set of ``Target Description'' keywords possibly related to such observations.\footnote{This was a somewhat cumbersome process, as the Target Descriptions are chosen by the individual PIs and have no checks or requirements for accuracy.} Our first search of the archive was completed on August 28, 2014 and resulted in a list of 480 candidate objects. We then used a SIMBAD search on the basis of the target coordinates to determine the type and redshift of these objects. Our initial selection of keywords yielded 96\% of AGNs/QSOs at $z<1.5$. The remaining objects consisted of higher redshift QSOs and a handful of stars from the Magellanic Clouds. 

We performed a follow-up search on April 20, 2016, which yielded a database of 527 objects, adding only 47 objects in the nearly 20-month span between the two searches. Therefore, we do not expect we would find a much larger sample if we were to search again. Removing stars and high redshift QSOs, our final database consists of 507 targets at $z<1.5$. We cross-correlated our database with HSLA when it was released and found the same list of objects. Although we adopted the HSLA spectra (but see \S\ref{s-comp-hsla} for some exceptions), we kept our original database information and structure for our survey. 

\subsection{Data reduction and Coaddition}\label{s-redux}
As we were finalizing our search for strong \hi-selected absorbers (\S\ref{s-search}), the HSLA became available, which provided coadded spectra across different programs of the same object. Since the HSLA data are likely to become the reference spectral database in the UV and since the data quality are equivalent to our initial database (see below), we decided to adopt the spectra from the HSLA for our survey. This is also beneficial for the reproducibility of our results. Furthermore the HSLA handles the errors properly, with a coaddition of the data in counts rather than flux that allows for a cleaner propagation of the photometric error \citep{gehrels86}. However, one of the limitations of the HSLA spectra is that they were obtained with no attempt to shift different exposures relative to one another to correct for any possible mismatch in the wavelength calibration. This can lead to blurred absorption line profiles, which could adversely affect our results. Since we did our own coaddition of the QSO spectra prior to the availability of the HSLA, we discuss below the cross-comparison between our original database and the HSLA after describing our coaddition procedure. 

\subsubsection{Original Coaddition}\label{s-original-coadd}
We used data reduced with the standard {\tt calcos} pipeline at MAST. While the  {\tt calcos}  pipeline provides a coaddition of the different exposures for a given program (without any cross-correlation of absorption lines), it does not coadd data across different programs or observed with different grating settings. In order to get the best SNR spectra, we needed to coadd data and we used the updated {\tt coadd\_x1d} algorithm from the COS GTO team (see \citealt{danforth16,keeney17}) to coadd the COS G130M and G160M spectra. We chose this algorithm because it maintains an appropriate flux calibration for the resulting spectra (critical for modeling the break at the Lyman limit if existent, see \S\ref{s-nhidet}), is fully automated, and has the possibility to coadd various settings straightforwardly. In view of the large sample that we analyzed, it was critical that little human interaction was required to produce the final datasets. 

This coadding program cross-correlates strong absorption lines to add the different exposures obtained from the same or different settings (different central wavelengths or gratings). More details can be found in \citet{danforth16}. The main caveat with this program is, however, that it does not handle the COS wavelength stretches that have been described in, e.g., \citet{wakker15}, where misalignments of a few tens of \km\ that vary as a function of wavelength can occur between different exposures. 

We tested the results from this coaddition with other coadditions that we undertook for our first survey \citepalias{lehner13} and our M31 halo survey  \citep{lehner15} where the individual exposures were coadded by two different algorithms. For the column densities, we systematically obtain comparable results for both the actual values of the column densities and the errors. The coadding program developed by \citet{wakker15} cross-correlates each line in each exposure over a small wavelength range to account for the COS wavelength stretches, providing currently the best wavelength calibration solution; this program was adopted for the M31 survey because accurate absolute velocities were critical. So, while the derived column densities agree typically within $1\sigma$ error, there were a few cases where the absolute velocities between the two reductions could differ by one or two COS resolution elements (i.e., 15--30 \km). As accurate velocities/redshifts are not too critical for the main aim of our survey (but see \S\ref{s-wavecal}), the fully automated {\tt coadd\_x1d} program provided coadded COS spectra of adequate quality. 

\subsubsection{Comparison with HSLA}\label{s-comp-hsla}
 We first visually compared our own and the HSLA coadditions, and overall the spectra look equivalent with similar SNR, except sometimes  in the G130M/G160M overlap region where the HSLA sometimes fails to correctly coadd the data. This can lead to an artificial break in the flux or much lower SNR in that region of the final spectrum (see in the current HSLA data-release, e.g., the spectra of PHL2525, 3C263, or MRK421). Generally, the overlap region between the two gratings was not critical for our work (i.e., in most cases no key absorption features or LLS breaks fell in that region). In cases where several transitions fell in that overlapping region, we adopted our initial coaddition (e.g., the absorber $z=0.556684$ toward J124511.25+335610.1 is such an example where several critical \hi\ transitions were present in that overlap region).
 
 Using the coadded spectra from our original database and the HSLA, we compared the normalized spectra for several absorbers as well as estimated column densities and kinematics. We found consistent results within less than $1\sigma$ (and, importantly, comparable errors). In particular, as detailed below, we re-analyzed several COS absorbers from the \citetalias{lehner13} sample and overall we derived column densities in agreement within about 1$\sigma$. 

Therefore, despite the inconvenience for a few absorbers where absorption feature fall in the G130M/G160M overlap region and because overall the HSLA and our coadded spectra gave consistent results, we decided to adopt the HSLA data. For ease of manipulating our database and plotting figures, we, however, enhanced the original FITS files provided by the HSLA by adding several keywords including the redshift of the objects, the right ascension, declination, Galactic longitude and latitude, the \hst\ and SIMBAD names, the minimum and maximum wavelengths, and the conversion needed to shift velocities to the Local Standard of Rest (useful, e.g., for aligning Milky Way absorption with \hi\ 21-cm emission profiles). 

\subsection{COS Wavelength Calibration}\label{s-wavecal}

As we already discussed in \S\ref{s-original-coadd}, there are known issues in the COS wavelength calibration, both in the absolute and relative wavelength calibrations. For the data used in our survey, we find that these effects are typically less than 20--30 \km\ as we now demonstrate.   

\begin{figure}[tbp]
\epsscale{1.2}
\plotone{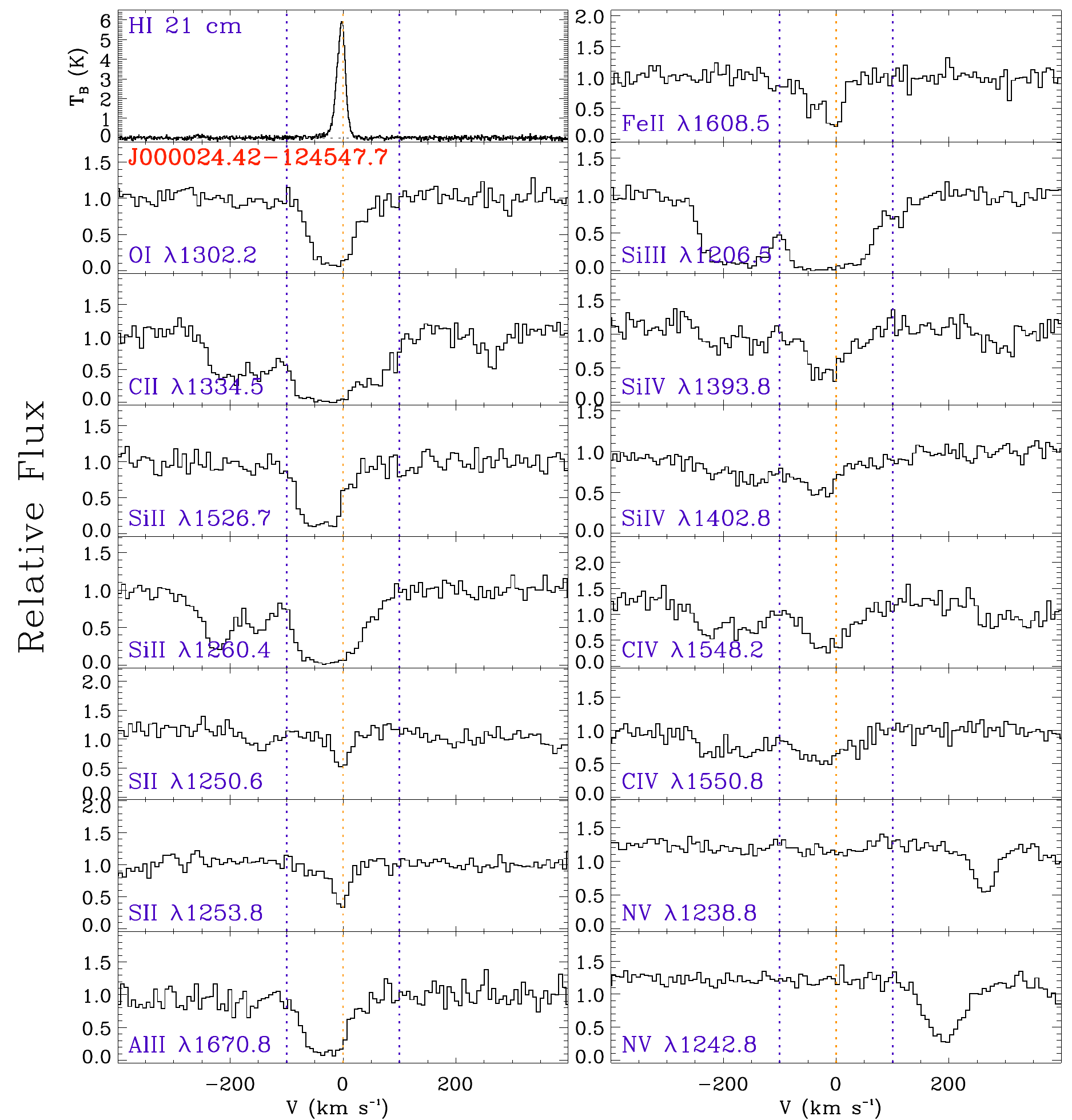}
    \caption{Velocity profiles for key atoms and ions for which the absorption arises from the Milky Way compared to the \hit\ emission ({\it top-left panel}) of the Milky Way toward the same direction. The vertical dotted lines just show  $-100, 0,$ and $100$ \km\ for reference. This type of figure was produced for each object in our database to help determining if there was any issue with the COS wavelength calibration. 
    \label{f-mw}}
\end{figure}

First, for all the 527 objects that were in our initial database, we visually compared the absorption seen in the atomic and ionic transitions from the Milky Way and the Galactic \hi\ 21-cm emission from the LAB survey \citep{kalberla05} observed in the same direction. As the \hi\ emission velocity is well calibrated, this comparison helps assessing the overall accuracy of the absolute wavelength calibration as well as possible relative shifts between different transitions if those originate from the same gas.  In Fig.~\ref{f-mw}, we show an example of a figure that we produced for all the objects in our database where we display a stack of velocity profiles of several atomic and ionic COS transitions and the \hi\ emission from the LAB data. Since \hi\ emission is probing the neutral gas, the best species to compare with is \oi\ or \nni, but these are often affected by strong airglow emission lines and strongly saturated. We therefore used singly ionized species and used also higher ions (\siiii, \siiv, \civ, \nv) to assess possible velocity differences between the ionized and neutral gas. In addition, we consider several transitions of each ion where possible because some Milky Way absorption features can be contaminated by IGM or CGM absorption at higher redshifts.  The velocity profiles shown in Fig.~\ref{f-mw} illustrate some of the issues. The saturated transitions of singly ionized species appear somewhat shifted relative to the main \hi\ emission, but they appear to be aligned with strong lines from doubly and triply ionized species. This is largely due to the strong saturation across several of the components in the low-velocity gas, as the velocity of the absorption in the weaker transitions of singly ionized species (e.g., \sii\ $\lambda$1253 and \feii\ $\lambda$1608) matches well the velocity of the main \hi\ emission. Hence for this target, we conclude that the absolute COS wavelength calibration is good to within one COS resolution element. We considered the calibration for our entire database in this way. We typically found comparable results, with absolute wavelength calibrations good to $\sim20$--30 \km. A notable exception in our survey is J154553.63+093620.5 where the absorption was shifted by $-50$ \km, which we subsequently corrected. 

Second, several targets were observed with ground-based telescopes to obtain \mgii\ $\lambda \lambda$2796, 2803  (see \S\ref{s-ground}). In that case, we can again test the relative and absolute wavelength calibration, but this time using the observed species at the redshift of the absorber. We found again that the velocity of \mgii\ (when detected) was typically aligned with the COS transitions in the redshifted restframe of the absorber. The main exception was again J154553.63+093620.5 where a similar velocity shift determined with the previous method was derived.

Third, we fitted  the \hi\ Lyman series with Voigt profiles for a large fraction of our sample. That is the only part in our survey where an alignment is critical to obtain the best possible fit. As we discuss in more detail in \S\ref{s-nhidet}, overall we found a good match between the central velocities of the different \hi\ transitions, suggesting the relative wavelength calibration was secure (to the limits we need) across at least the wavelengths probed by the Lyman series absorption. Only in a few rare cases did we need to shift a specific \hi\ transition to match the velocities of the other \hi\ lines in order to have a better-fitted profile (this was automatically taken care of in the profile fitting, see \S\ref{s-nhidet}).

\subsection{Adopted Database}\label{s-adopt-data}

While our initial database had 527 QSO spectra, the final sample for our survey was reduced to 335 QSOs based on quality cuts to those observations and selection criteria of our targets. Deriving accurate \nhi\ for absorbers with $\mlnhi \ga 15.2$ requires 1) observability of several Lyman series transitions beyond \lya\ and \lyb\ (i.e., $z_{\rm em} \ga 0.17$ to redshift the Lyman series in the COS wavelength) and 2) SNR\,$\ga 1$ in the COS spectra with a bin size of $\sim$6 \km\  (or SNR\,$\ga 2$ per resolution element --- hereafter any SNR value is for a COS spectrum with 6 \km\ bin size; we searched for absorbers in even lower SNR spectra, but none were reliably found; see \S\ref{s-search} for more details).\footnote{Ultimately all the QSO spectra with absorbers included in our \hit-selected survey have ${\rm SNR}\ga 2.5$ (there are 294 QSOs satisfying this SNR threshold). In the 41 QSO spectra with  $1< {\rm SNR}< 2.5$, we found only two very strong \hit\ absorbers for which we could derive only a lower limit on \nhi.} 

In Table~\ref{t-general}, we summarize the QSO sample over which we carried out our search for strong \hi\ absorption, sorted by increasing right ascension. The first column gives the J2000 name, constructed using the right ascension and declination in J2000 sexagesimal format (JHHMMSS.ss+DDMMSS.s; note that the output is truncated in precision rather than rounded following the IAU recommendation and the official SDSS designation). The second column gives the PI-provided name from the \hst\ observations, i.e., the name as displayed in MAST. We provide that information to ease the search in the MAST archive (many of the ``\hst\ names'' are not standard and thus will not be resolvable in SIMBAD).  The third column gives the redshifts that we extracted from SIMBAD (overall these are equivalent to those listed in the HSLA that come from NASA/IPAC Extragalactic Database --- NED). The next three columns give the wavelength coverage and median SNR of the COS spectra. Spectra with $\lambda_{\rm max} \la 1480$ \AA\ were obtained with the COS G130M grating setting only; spectra with $\lambda_{\rm min} \ga 1400$ \AA\ were obtained with the G160M grating setting only; spectra that cover wavelengths from $\sim 1130$ to $1800$ \AA\ were obtained with both COS G130M and G160M gratings. The seventh column provides the \hst\ program identification number of the original program(s) used for each target. The eighth column provides information regarding the number of absorbers with $\mlnhi >15.1$ found in a given spectrum; a number between parenthesis means that a strong \hi\ absorber with $\mlnhi \ga 17.9$ was found but we were not able to constrain \nhi\ better than a lower limit. Finally, the last column provides information if supplemental ground-based spectra exist with the name of the instrument used. These ground-based spectra cover, in particular, the strong \mgii\ $\lambda \lambda$2796, 2803 doublet discussed below. 

\subsection{Supporting Ground-Based Observations}\label{s-ground}

Our survey includes new and archival spectroscopy obtained with ground-based instruments. As demonstrated in \citetalias{lehner13} and \citetalias{wotta16},  the \mgii\ $\lambda \lambda$2796, 2803 doublet is an excellent metallicity tracer for pLLSs and LLSs (and, as we will show in paper III, even for the weaker absorbers with $15.2 \la \mlnhi \la 16.1$) because it is so strong. A large fraction of the ground-based observations also covers the strong transitions of \feii\ $\lambda \lambda$2382, 2600 (depending on the redshift of the absorbers and wavelength coverage in the blue), providing additional information on nucleosynthesis and depletion effects (which are mild; see \citetalias{lehner13} and \S\ref{s-res-ratios}). 

The two main ground-based telescopes and instruments used in our survey are the High Resolution Echelle Spectrometer (HIRES, \citealt{vogt94}) on Keck I and the Ultraviolet and Visual Echelle Spectrograph (UVES) on the Very Large Telescope (VLT) \citep{dekker00}. For one absorbers we also used low resolution spectra from SDSS and  for another one the Multi-Object Double Spectrograph (MODS) on the Large Binocular Telescope (LBT); we note that the absorption for these two absorbers is very strong and only lower limits could be derived on the column densities of \mgii. For the LBT/MODS data processing, we refer the reader to \citetalias{wotta16}. 

For the UVES data, archival data was used for 4 sightlines, retrieved from the ESO archive. Another \uvesobs\ targets were observed through our ESO VLT/UVES program 0100.A-0483, which was awarded up to 31 hours of VLT/UVES ``filler" time during period 100A (data were acquired from September 2017 to March 2018). This program used a 1\arcsec\ slit with Dichroic \#1 to obtain $R\sim40,000$ data target \mgii\ absorption from CGM absorbers. For sightlines with systems below $z \la 0.25$, we used the 346+564 grating setting to provide coverage over the wavelength range $3200 \la \lambda \la 3900$ and $4700 \la \lambda \la 6600$ \AA. For all others we used the 390+564 setting, producing usable data over roughly $3300 \la \lambda \la 6600$ \AA. The nominal exposure times were designed to achieve SNR\,$\ga 10$ per pixel for \mgii\ at the lowest-redshift absorber along a sightline. Our observations were taken over a wide range of conditions, and not all observing setups for a given target were necessarily executed. Thus, the SNR achieved varies between targets (as well as with wavelength for a given target). To reduce the UVES spectra (archival and new data), we use the ESO pipeline data reductions, which is adequate for the data quality that we obtained. 

The bulk of our ground-based observations come from Keck HIRES observation. In total 53 sightlines in our sample were observed with Keck HIRES. About 75\% of this sample comes from the KODIAQ database available at the Keck Observatory archive (KOA) \citep{omeara17,omeara15,lehner14}. KODIAQ consists of a uniformly- and fully-reduced sample of QSOs by our group at $0.07 < z_{\rm em} <  5.29$ observed with Keck HIRES at high resolution. Most of the spectra for QSOs with $z\la 1.3$ in the KODIAQ database were initially obtained to support two Large \hst\ programs, COS-Halos and the COS Absorption Survey of Baryon Harbors (CASBaH) \citep{tumlinson11a,tripp11}. The remaining 25\% were obtained through the NASA and University of Hawai`i at Hilo Keck allocation time (PIs: O'Meara and Cooksey, respectively). We applied the same data reduction to these data that was applied to the KODIAQ sample \citep{omeara15}, and these spectra will be included in the next KODIAQ data release. 

In total, there are 64 high-resolution HIRES and UVES QSO spectra that have coverage of \gbmgii\ absorbers (see Table~\ref{t-general} for the listing of sightlines observed from the ground). The absorbers toward J113327.78+032719.1 and J213135.20-120704.5, observed with HIRES and UVES, respectively, both had very strong absorbers with $\mlnhi \ga 18$; as only a lower limit could be derived on the \nhi, these two absorbers are not included in our final sample for studying metallicities. In \S\ref{s-res-col}, we provide more information regarding the sensitivity of the ground-based observations, but our general aim was to obtain spectra with SNRs that gave $W_\lambda \la 25 $ m\AA\ at the $2\sigma$ level (over a typical integration range of $\pm 15$ \km\ or a full velocity width of about 30 \km). This sensitivity level is critical to placing robust limits on low-metallicity gas. 

\section{Search for Strong \hi-selected Absorbers}\label{s-search}

Our survey is based on a search of the entire database of COS G130M and/or G160M spectra for absorbers with $\mlnhi \ga 15.2$. Absorbers with $\mlnhi \ga 16.5$ create a strong-enough break at the Lyman limit in the continuum of the QSO so that a visual inspection of the spectra allows one to identify these absorbers readily, even in low SNR spectra.\footnote{In the higher SNR spectrum (as those analyzed in the \citetalias{lehner13} sample), a break can be observed down to $\mlnhi \ga 16.2$, which was the \hit\ column density limit in \citetalias{lehner13}.} However, for lower \nhi\ absorbers, the signature of the Lyman break and even the convergence of the Lyman series near the Lyman limit are not readily observable or more difficult to discern. 

\begin{figure}[tbp]
\epsscale{1.2}
\plotone{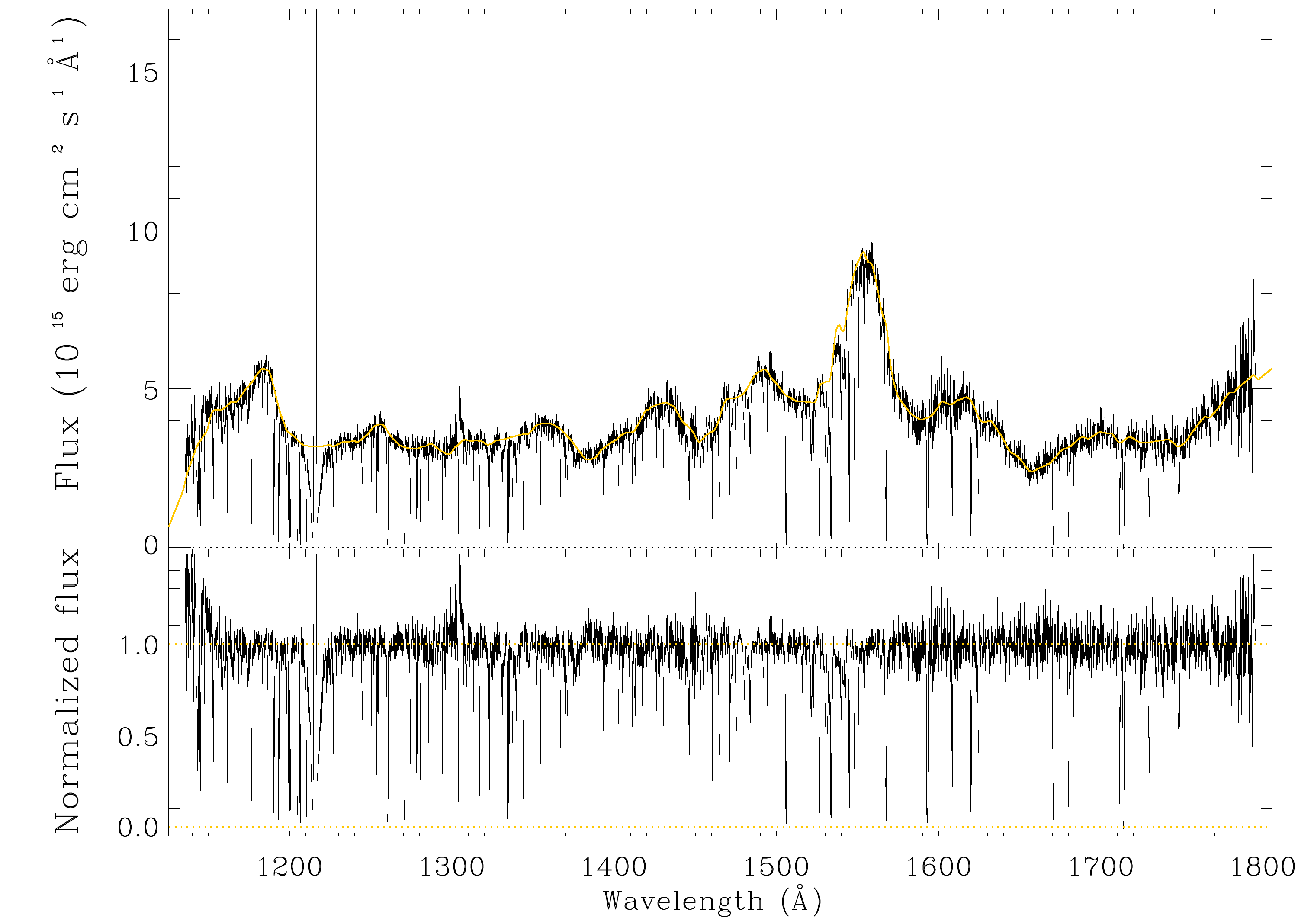}
    \caption{Example of a continuum fit on the entire observed spectrum ({\it top}) and resulting normalized spectrum ({\it bottom}) for the QSO J110539.79+342534.3 at $z_{\rm em} = 0.50901$. This continuum was solely used for the automated search of strong \hit\ absorbers in each QSO spectrum in our database.
    \label{f-example_spectrum}}
\end{figure}

We therefore developed and used an automated search tool for identifying strong \hi\ absorption. The initial step in the search was to fit a continuum to the entire spectrum of a given QSO. While this large-scale continuum is not adequate for quantitative measurements of the absorption lines, it is good enough to allow a search for strong \hi\ Lyman series lines. To enable this continuum fit, we first masked all the relatively strong absorption features observed in the QSO spectrum. The continuum estimate was then a heavily rebinned version of the input spectrum after masking. We interpolated across the masked regions assuming little variation in the continuum level over these masked regions. Since abrupt changes in the continuum slope can be observed near the peak of emission lines, we allowed for a refined binning over such regions. An example of the final continuum model is shown in Fig.~\ref{f-example_spectrum}, which demonstrates that this approach provides a continuum model good enough to enable a search for specific absorption features in the spectrum. We emphasize that we did not use this continuum model to make quantitative measurements of the absorption lines.

We identified strong \hi\ absorption systems based on the presence of Lyman series transitions from 1215.7 \AA\ (\lya) down to 937.8 \AA\ (Ly$\epsilon$). Absorbers with $\mlnhi \ga 15.2$ should produce visible absorption down to at least Ly$\epsilon$, and we used this as a criterion for identification of such absorbers. At the resolution of COS G130M and G160M data, \hi\ absorbers in our target range always show saturated absorption in \lya\ and \lyb. We therefore also required that if \lya\ and/or \lyb\ are covered for a candidate absorber, at minimum three contiguous pixels in the cores of these lines reach a value at or below 0.5 times the flux value corresponding to the $1\sigma$ error on the flux. These search criteria are insensitive to the broadening of the absorbers as long as the Doppler parameter $b\la 80$ \km. Using our profile fitting code (see \S\ref{s-nhidet}), we generated absorption profiles convolved with the COS G130M and G160M line-spread functions for an absorber with $\mlnhi = 15.3$ from \lya\ to Ly$\epsilon$ where $b$ was allowed to vary from 20 to 80 \km\ and overplotted these profiles on a COS G130M and G160M spectrum with SNR\,$\simeq 5$, showing that even if $b=80$ \km, absorption would still be detected in \hi\ $\lambda$$\lambda$937.7, 949.7 transitions and the profiles of \lya\ and \lyb\ would be strongly saturated.\footnote{We did not consider broader components since \citet{lehner07} show these are quite rare and are only seen for \hit\ absorbers with $\mlnhi \la 13.4$. In fact, we note that according to their high resolution (\hst/STIS E140M) survey, the 8 absorbers with $\mlnhi \ga 15.1$ have all $b \la 50$ \km\ in their survey (see Fig.~5 in \citealt{lehner07}).} We are therefore confident that our search algorithm did not miss broad \hi\ absorbers with $\mlnhi \ga 15.2$. As we will show in \S\ref{s-res-vel} there is no evidence of absorbers with $b> 50$ \km\ and $\mlnhi \ga 15.2$ at $z\la 1$, consistent with the smaller but higher resolution survey by \citet{lehner07}.

Our program computed the wavelength path to search based on the combination of the  maximum and minimum observed wavelengths and redshift of the QSO. Since we require access to transitions weaker than \lyg\ to derive accurate \nhi\ values, the lowest redshift in our survey when G130M data are available is $z_{\rm min}\sim 0.2$. The maximum redshift accessible with the G160M data is $z_{\rm max}\sim 0.9$. Our search program scanned each QSO spectrum starting from the highest redshift available (either observable or the QSO redshift); checked for the existence of Lyman series absorption from \lya\ to Ly$\epsilon$ (if \lya\ is outside the wavelength range, it adopted \lyb, or \lyg, etc.); and, if the  \lya\ or \lyb\ transitions are covered, ensured that several contiguous pixels in the line centers were saturated (see above), helping to reduce the number of false positive candidates. If all these conditions were satisfied, the program produced a stack of normalized profiles against the rest-frame velocity from \lya\ to Ly-11 $\lambda918.12$ at the identified redshift of the absorber. The program then iterated to search for additional potential \hi\ absorber(s) in a given QSO spectrum. Each stack of velocity profiles was then visually inspected to eliminate any residual false-positive absorbers (e.g., absorbers with $\mlnhi \la 15$). To facilitate this examination, 3 \hi\ models (convolved with the COS LSF) were over-plotted for each transition showing the absorption expected for $\mlnhi = 15$, 16, and 17 and $b=30$ \km.

We also independently performed a visual search for the signature of a break (or multiple breaks) caused by Lyman limit absorption in each QSO spectrum and determined the redshift of those using the Lyman series lines. The two searches led to the same identification  of the strong systems with $\mlnhi \ga 16.5$, with the following exception: where the automatic search program failed to identify an absorber that was identified visually, it was because the Lyman series transitions above 937 \AA\ were not covered. None of these absorbers ended up in our sample because we could not reliably determine \nhi\ for any of these absorbers (i.e., the Lyman limit break was saturated with  no flux recovery). 

Two recent surveys were undertaken with a smaller sample of COS spectra by \citet{stevans14} and \citet{shull17}. We emphasize that we did not use their results for our survey (neither their column densities, nor their identification of the absorbers), and thus they serve as a good check on our search methodology. We compared the results of our survey with the absorbers reported in \citet{shull17}, who focus on absorbers with $\mlnhi \ga 16$. All of the systems found by \citeauthor{shull17}\ were identified in our search. The survey conducted by \citet{stevans14} included absorbers with $\mlnhi \ga 13.4$. We compared the results for a few sightlines. Our search identified all of their absorbers meeting our selection criteria (redshift $\mzabs \la 0.9$  and $\mlnhi \ga 15.2$), but absorbers for which \nhi\ could not be accurately estimated (to better than $\sim 0.3$ dex in \lnhi) did not make it into our final sample.

The redshift of each absorber was determined from the peak optical depth of the strongest \hi\ component in the automatic search process.  In a few cases, this process was affected by contamination from other lines. In these cases, we adjusted the redshifts if they were offset by more than $\sim$1 COS G130M--G160M resolution element ($\sim\pm15$ \km). When they are detected, we find that low metal ions such as \cii, \oii, and \mgii\ are at the same redshifts as the strongest \hi\ absorption  component for a given absorber (within the COS velocity error of about 15 \km).

\section{Estimation of the Column Densities}\label{s-estimate-col}

Our automated search provided a sample of \ssz\ absorbers. For each of the absorbers, we estimated column densities for all covered (and uncontaminated) transitions given in Table~\ref{t-ion-list}, amounting to column density estimates for several thousands of absorption lines. We developed a semi-automated approach to measuring these column densities in order to expedite and systematize the process. For each transition, an automated continuum was estimated as described in \S\ref{s-cont}. Once the continuum placement was set, we employed the apparent optical depth (AOD) method \citep{savage91} as described in \S\ref{s-abs} to estimate the total ionic or atomic column density. For \hi, we used a combination of different methods as described in \S\ref{s-nhidet} depending in part on the strength of the absorption. Throughout we assumed the atomic parameters for the UV  and EUV lines listed in \citet{morton03} and \citet{verner94}, respectively, including central wavelengths and $f$-values. 

\subsection{Continuum Modeling}\label{s-cont}

\begin{figure}[tbp]
\epsscale{1.2}
\plotone{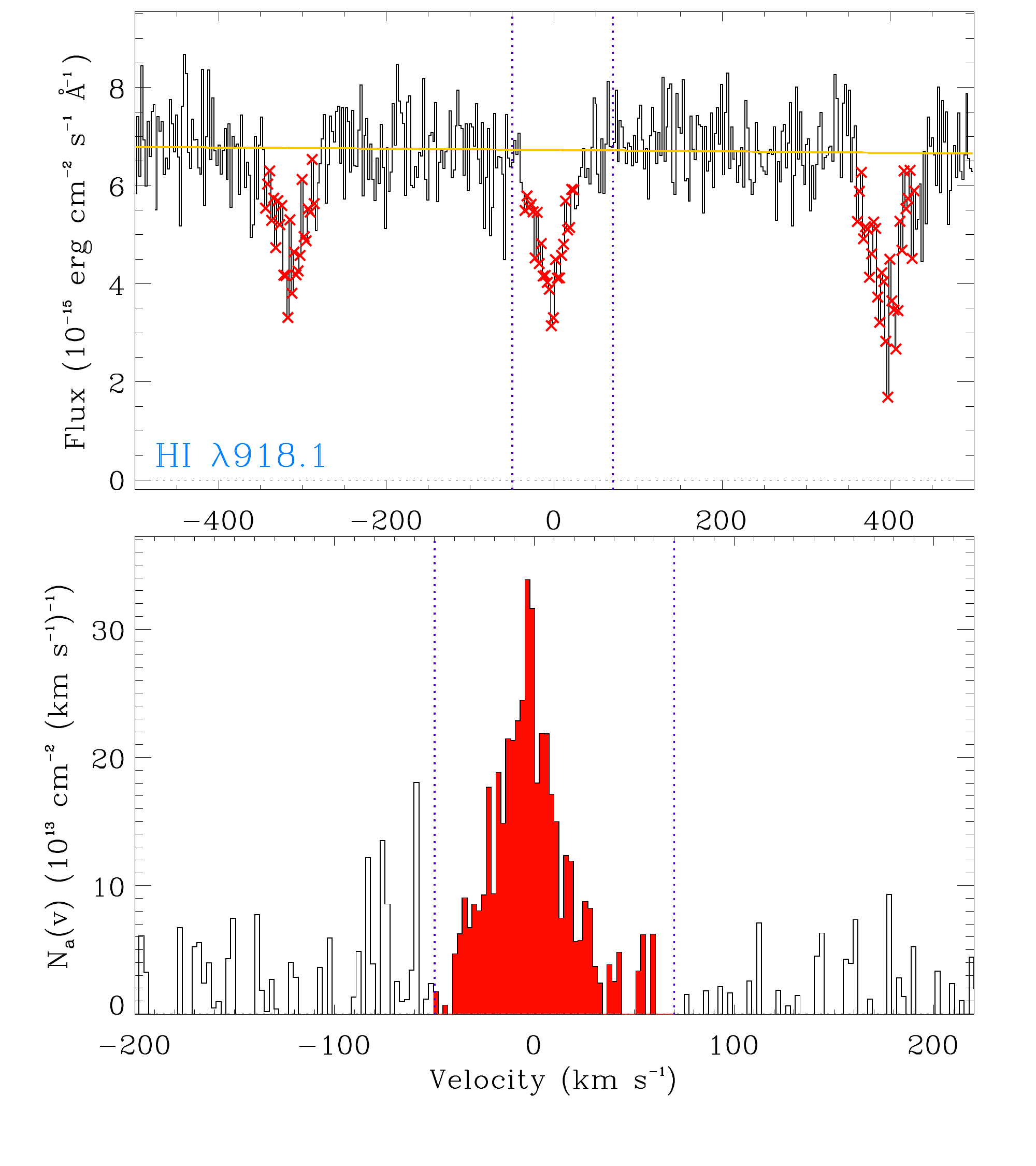}
    \caption{Example of a localized automated continuum fit ({\it top}) and apparent column density profile ({\it bottom}) of the \hit\ $\lambda$918 transition observed at $z=0.347925$ toward J000559.23+160948.9 (note that top and bottom $x$-axis scales are not the same). On the top panel, the red crosses mark points rejected during the continuum fit; the solid yellow line shows the Legendre polynomial fit to the continuum near the line of interest. The vertical dotted lines mark the integration range, while in the bottom panel, the red area marked the integrated column density profile. This type of figure was realized for each transition listed in Table~\ref{t-ion-list} and was visually inspected to ensure that the continuum model was satisfactory.  
    \label{f-example_cont}}
\end{figure}

To determine the line properties (equivalent widths, column densities, and velocities), we automatically fit the continuum near the absorption features using Legendre polynomials. A velocity region of about $\pm$2000 \km\ around the relevant absorption transition  was initially considered for the continuum fit. This velocity interval could vary as described below, depending on the complexity of the spectrum (e.g., a region of a spectrum with a high density of absorption lines or with QSO emission lines). In all cases the interval for continuum fitting was never larger than $\pm$2000 \km\ or smaller than $\pm$250 \km. Within this pre-defined region, the spectrum was broken into smaller sub-sections and then rebinned. The continuum was fit to all pixels that did not deviate by more than $2\sigma$ from the median flux in their local subsection. This removes pixels from the fitting process that may be associated with small-scale absorption or emission lines. Legendre polynomials of orders between 1 and 5 were fit to the remaining (non-rejected) pixels, with the goodness of the fit determining the adopted polynomial order. (Typically the adopted polynomials were of orders between 1 and 5 owing to the relative simplicity of the QSO continua when examined over velocity regions as small as 500--4000 \km). After a few tests, we realized that in several cases the continuum was not quite adequate near some absorption lines, in particular not rejecting completely the wings of the absorption lines. A second pass was necessary where the data were binned again and any binned pixels that were above or below 1.7$\sigma$ below the original Legendre fit were rejected from the continuum placement regions. After several trials with different clipping values, we determined that the 1.7$\sigma$ value allowed the algorithm to reject the highest number of non-continuum points, which includes unrelated absorbers and, most importantly, the tails of the absorption lines.  This last part was crucial to getting a good fit to the continuum near the primary absorber itself. Fig.~\ref{f-example_cont} shows an example of the continuum fit around the \hi\ $\lambda$918 line observed at $z=0.347925$ toward J000559.23+160948.9.

This procedure was applied to a pre-defined set of transitions (see Table~\ref{t-ion-list}), with the continuum defined locally for each. Each continuum model was visually inspected to ensure it was satisfactory. In a few cases, the automatic continuum fitting failed owing to a complex continuum (e.g., near the peak of an emission line or where many absorption lines were present within the pre-defined continuum window). In these cases, we first tried to adjust the velocity interval of the spectrum to provide better-constrained fits; if that still failed, we manually selected the continuum region to be fit. 

\subsection{Absorption Line Properties}\label{s-abs}

The next step of the analysis was to integrate the absorption lines to determine the kinematic properties (central average velocities and line-widths), equivalent widths, and column densities for each absorption feature. The velocity range over which the line was integrated was determined from the \hi\ absorption and, as much as possible, the integration corresponded to a single absorbing complex (at the COS resolution).  

To estimate the column density, we use the apparent optical depth (AOD) method. In this method, the absorption profiles are converted into apparent optical depth per unit velocity, $\tau_a(v) = \ln[F_{\rm c}(v)/F_{\rm obs}(v)]$,  where $F_c(v)$ and $F_{\rm obs}(v)$ are the modeled continuum and observed fluxes as a function of velocity.  The AOD, $\tau_a(v)$, is related to the apparent column density per unit velocity, $N_a(v)$, through the relation $N_a(v) = 3.768 \times 10^{14} \tau_a(v)/(f \lambda(\mbox{\AA})$)  ${\rm cm}^{-2}\,({\rm km\,s^{-1}})^{-1}$, where $f$ is the oscillator strength of the transition and $\lambda$ is the wavelength in \AA. The total column density is obtained by integrating the profile over the pre-defined velocity interval, $N = \int_{v_{\rm min}}^{v_{\rm max}} N_a(v) dv $, where  $[v_{\rm min},v_{\rm max}]$ are the boundaries of the absorption. We computed the  average line centroids through the first moment of the AOD $v_a = \int v \tau_a(v) dv/\int \tau_a(v)dv$ \km. 

We characterize the total velocity widths of the absorption profiles by measuring the width \dv\ containing the central 90\% of the integrated AOD in the line \citep[i.e., \dv\ is the difference in velocity between the pixels at which 5\% and 95\% of the total optical depth has been reached;][]{prochaska97,fox13}. For absorbers with several transitions of a given species detected at more than $2\sigma$, we took a robust mean of the measurements, rejecting outliers more than $3\sigma$ from the median value (this is mainly useful for the \hi\ transitions where weak and strong lines were compared). 

We integrated the equivalent widths using the same  $[v_{\rm min},v_{\rm max}]$ integration range adopted for the $N_a(v)$ profiles. The main uses of the equivalent widths in our survey are twofold: 1) to determine if the absorption was detected at the $\ge 2\sigma$ level, and 2) to construct a curve-of-growth model for the \hi\ absorption (see below).  In cases where the line was detected at $\ge 2\sigma$ significance, the estimated apparent column density is adopted. Otherwise, we quote a 2$\sigma$ upper limit on the column density which is defined as twice the 1$\sigma$ error derived for the column density assuming the absorption line lies on the linear part of the curve of growth.

When an absorption feature is detected at $\ge 2\sigma$ significance, the AOD measurement may not represent the best estimate of the true column density, since it does not yet include an assessment of line contamination or saturation. We decided early in our survey that to embark into a complete line identification was not feasible owing to the large size sample and more importantly the non-complete wavelength coverage for many QSO spectra. Many spectra have gaps in the wavelength coverage and have only G130M or G160M observations. We assessed both line contamination and saturation using procedures described in \citetalias{lehner13} and \citetalias{wotta16}. Below we describe the key steps and assumptions we made to determine the possible levels of contamination and saturation of an absorption feature.\footnote{Our goal is to determine the metallicity of the weakly ionized gas that is observed over the same velocity interval as the strongest \hit\ components. This requires we model accurately the properties of at least the singly and doubly ionized species, if not those of the higher ions (such as, e.g., \siivt), depending on the \hit\ column density of the absorber. Hence, although we have determined and report the column densities for higher ions (e.g., \oivt, \ovit), it is quite possible that our measurements do not represent an integration of the entire absorption profiles, since we only consider the absorption seen in the main \hit\ absorption (i.e., the component that gives rise to at least one absorber with $\mlnhi \ga 15.2$). One can refer to the figures in the Appendix to check the integration range for each reported ion. We caution the reader not to use blindly the column densities of the high ions reported in this work, which were derived for a very specific purpose, for issues beyond the metallicity of the main absorption complex. We will revisit the high-ion properties, and in particular \ovit, in paper IV.}

Our strategy for assessing whether a given transition is saturated or contaminated by another, unrelated absorber is based on a combination of visual inspection and comparison of the relevant quantities (e.g., mean velocity and apparent column density). Strong contamination can often be diagnosed visually, comparing the velocity structure of the absorption profiles across transitions from the same or similar ions (to determine if the main part of the absorption is at similar velocities or not). Our measurements of the mean velocities from the integration of the absorption profiles can often point to strong contamination (if the mean velocity from one transition is significantly different than others that are aligned). Typically, we allow for 10--15 \km\ uncertainty in the velocity in that comparison (owing to stretch in the COS wavelength solution). In the case of obvious contamination, the absorption feature is discarded. The only exception is if the contamination is very mild, only occurring in the wing of a relatively strong absorption; in that case we changed slightly the velocity integration range to avoid the contaminated portion for that absorption feature. Similarly strong saturation where the apparent absorption reaches zero flux and the velocity profile can be reliably assessed visually (we made sure that there was no obvious sign of contamination in that case).

For less obvious contamination or saturation levels, the procedure depends on the number of transitions for a given ion or atom. For \hi, both the saturation and contamination effects can be assessed reliably since several transitions are always available and several independent methods are used to determine the column density (AOD, COG, profile fitting, and/or break at the Lyman limit, see \S\ref{s-nhidet}). Similarly some metal ions (e.g., \oii, \oiii, \siii) have several transitions with large enough $\lambda f$-values to allow us to directly determine if there is some evidence for saturation or contamination. For atomic or ionic transitions that have $\Delta (\lambda f) \simeq 2$ and the difference in apparent column densities of the weak and strong transitions is $\Delta (\log N_a) \equiv \log N^{\rm weak}_a - \log N^{\rm strong}_a \le 0.13$ dex \citepalias[see][]{wotta16}, we were able to correct for the mild saturation using the method described in \citet{savage91} and \citetalias{wotta16}. In this case, we report the apparent column densities for each transition and then the adopted column density, corrected for saturation. In cases where we were not able to correct for saturation, we report the column density as a lower limit determined by the AOD method and increased by 0.15 dex, which corresponds to the maximum saturation correction we can apply reliably (see \citetalias{wotta16}). 

To check for contamination in single transitions (in particular \ciii\ and \siiii\ that are critical for constraining the ionization models), we first systematically checked for strong Milky Way contamination by comparing the observed wavelengths of the ions from  the absorbers to the wavelengths of known Milky Way absorption lines (e.g., \cii, \siii, \siiii, \nni, \oi, \feii). Second, we used the information provided by atoms and ions with several transitions  and the relative abundances. For the latter, while ionization obviously plays a role, one can deduce straightforwardly if absorption features are contaminated (e.g., if $N_{\rm NI} \ge N_{\rm CII}$ or $N_{\rm NII} \ge N_{\rm CIII}$, then \nni\ or \nii\ are very likely contaminated, respectively). Using the information from these multiple transitions, we can also assess for a given absorber at what peak optical depth there is evidence for saturation; if there is for an ion or atom with only one available transition, the column density is reported as lower limit with the 0.15 dex correction (see above).

The column densities, velocity integration ranges, averages velocities, and redshifts are summarized in Tables~\ref{t-results} and \ref{t-results-l13}, which correspond to the new sample with $15.1 \la \mlnhi <19$ and the COS G130M and G160M absorbers from the \citetalias{lehner13} sample that were re-analyzed here (including any new ground-based spectra taken for this work), respectively. We also provide all the results in a machine readable format (see Appendix). 

\subsection{\hi\ Column Densities}\label{s-nhidet}
To estimate the \hi\ column density, we used several methods, which in part depend on the strength of the absorbers. For absorbers with $\mlnhi \ga 16.8$ and where the \hi\ column density is clearly dominated by a single component or where we cannot reliably estimate the column densities in the individual components of \hi\ {\it and}\ metal ions, we used the flux decrement at the Lyman limit. 
For this method, we followed the procedure described in \citetalias{wotta16}. In short, we estimated its redshift from the Lyman series lines. We then fitted the continuum on the red side of the break with a composite QSO spectrum \citep{telfer02} and applied an artificial flux decrement at the wavelength of the break. We adjusted the modeled $\tau_{\rm LL}$ to match the flux decrement (and recovery when possible) on the blue side of the break, producing an estimate for \nhi. The high-resolution data have high enough SNRs that we could confidently fit the break at the Lyman limit with little difficulty (given coverage of the continuum and recovery). The error on \nhi\  is typically $\sim$0.05--0.15 dex using this method for $\mlnhi \ga 16.8$. In several cases, we also fitted the Lyman series lines with Voigt profiles (see below), and systematically found consistent results. The \hi\ column densities derived using the break method have a LL line entry in Tables~\ref{t-results} and \ref{t-results-l13}.

\begin{figure}[tbp]
\epsscale{1.2}
\plotone{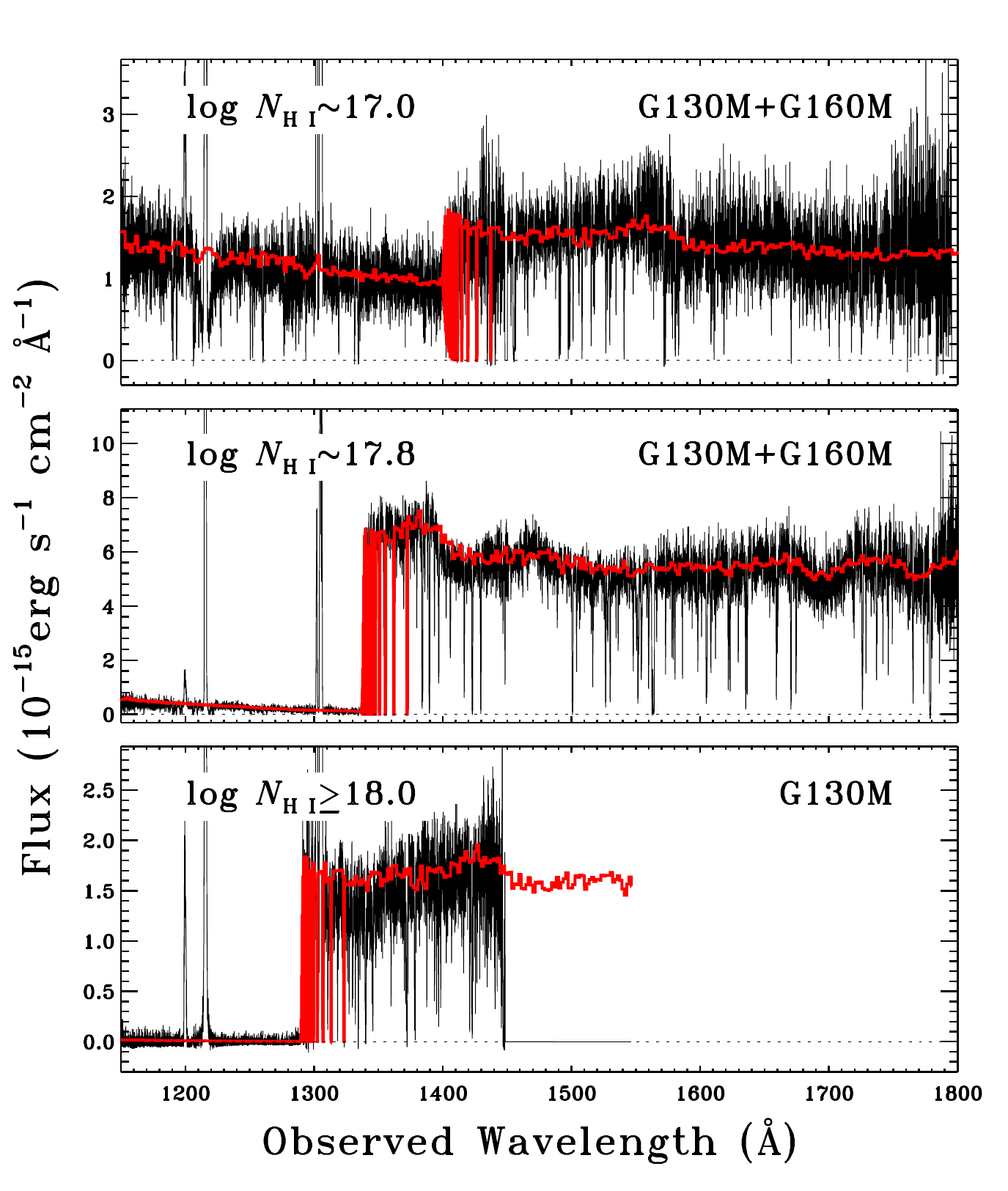}
    \caption{Examples of high-resolution spectra taken with COS G130M and G160M from our CCC survey. We show moderate- ({\it top}), high- ({\it middle}), and very high-\hit\ column density (saturated, in this case; {\it bottom}) absorption. \label{f-strongLLS_lyman_breaks}}
\end{figure}

Depending on the complexity of the QSO continua and interstellar spectra (in particular near the Milky Way \lya\ absorption), the method from the break at the Lyman limit can be used to determine \nhi\ for absorbers with  $\mlnhi \ga 16.3$ or $ \tau_{\rm LL}\ga 0.15$. However, for the typical SNR of the COS spectra (which is lower on average than the L13 sample), the break technique produces less accurate \nhi\ than the other methods described below that use the Lyman series lines. We therefore only used this method as a verification step for absorbers with $16.3 \la \mlnhi \la 16.9$. In Fig.~\ref{f-strongLLS_lyman_breaks}, we show examples of this method for absorbers with different \hi\ column densities. We overplot in red a composite QSO spectrum from \citet{telfer02} to which we have applied an artificial break at the adopted \nhi\ as described above. The recovery at the blue end of each spectrum is useful for constraining the \nhi\ using the break, even when the continuum absorption and Lyman series lines are saturated near the break. However, at the highest column densities, the recovery is no longer present (see the bottom panel of Fig.~\ref{f-strongLLS_lyman_breaks}) and we could only derive a lower limit on \nhi. Therefore, we estimated the \hi\ column densities of absorbers using the decrement method only for absorbers with $16.9 \la \mlnhi \la 18$ (the exact value at the high end depends on how much flux recovery is available). 

For absorbers with  $15.2 \la \mlnhi \la 16.9$, we used a combination of different techniques to estimate \nhi, all using the Lyman series transitions: the AOD method, profile fitting (PF), and curve-of-growth (COG). For the AOD method, we used all the available weak Lyman transitions that are not contaminated and for which the continuum can be reliably modeled. Typically for the  weak Lyman transitions, the  saturation effects are mild in the $15.2 \la \mlnhi \la 16.8$ range, but for the stronger absorbers or in cases where the weakest transitions are not available (owing to a lack of wavelength coverage or contamination), the COG and PF methods were favored to determine \nhi. In nearly all  cases, we determined \nhi\ with at least two transitions when we used the AOD; these transitions are summarized in Tables~\ref{t-results} and \ref{t-results-l13} (in the rare cases where we did not, the adopted \nhi\ value is based only on the COG and PF methods). If there is no evidence of saturation, we averaged the values and propagated the errors accordingly, which is given in the \hi\ AOD line entry in Tables~\ref{t-results} and \ref{t-results-l13}. In the rare cases, where there is evidence for saturation, we applied the method described in \S\ref{s-abs} to correct for it. 

The COG method used a $\chi^2$ minimization and $\chi^2$ error derivation approach outlined by \citet{sembach92}. A single component COG is assumed. The program solves for $\log N$ and $b$ independently in estimating the errors. We started with all the \hi\ transitions for which we estimated the equivalent widths and did not appear {\it a priori} to be contaminated based on the AOD analysis. We checked the results from the COG model and then removed only the transitions clearly departing from the model. We ran our COG program again with the new set of transitions to obtain the final estimate of \nhi\ with the COG method. Our adopted COG values are summarized in Table~\ref{t-results} in the COG line entry. 

\begin{figure}[tbp]
\epsscale{1.1}
\plotone{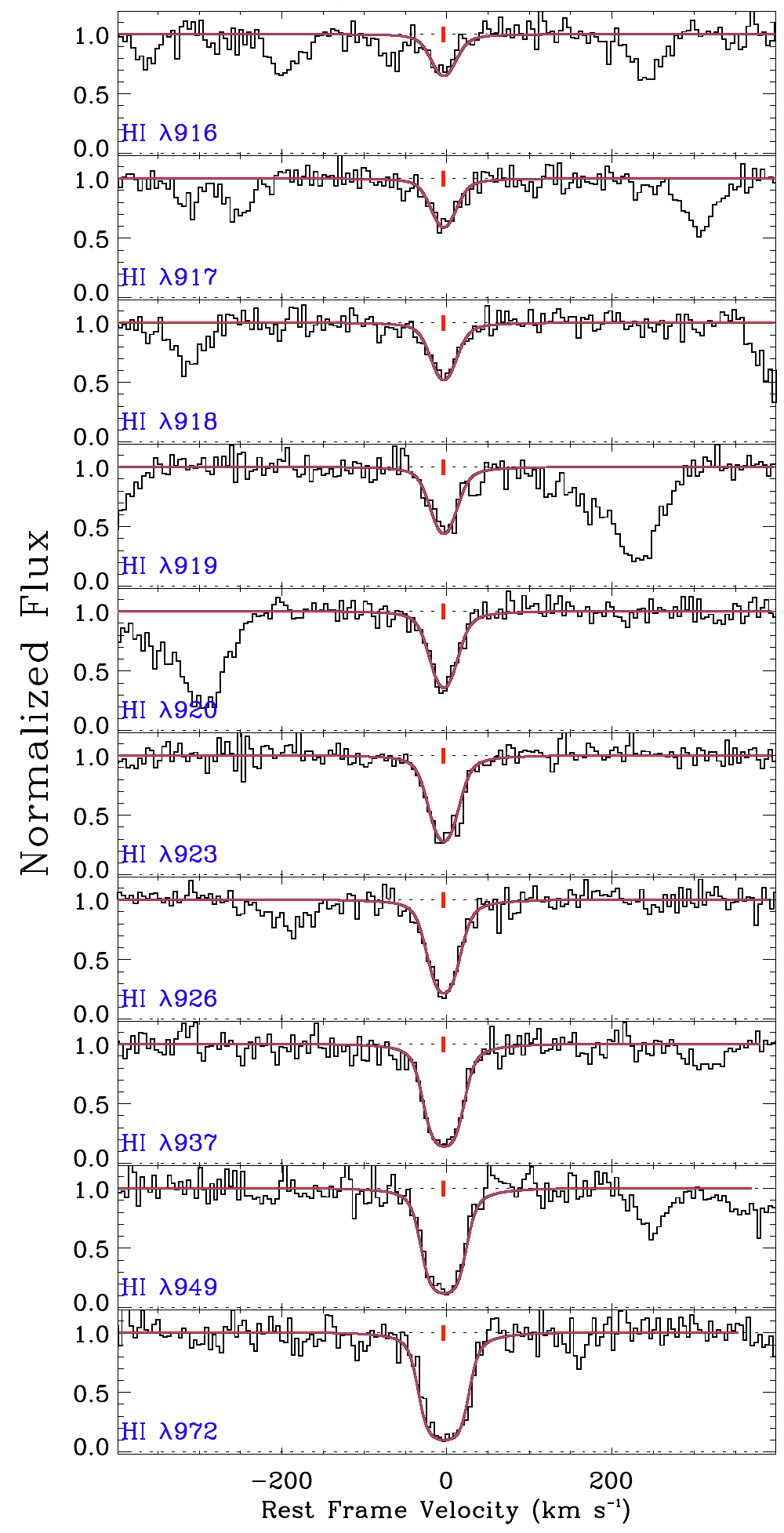}
    \caption{Example of a pLLS observed at $z=0.347925$ toward J000559.23+160948.9 where we fitted the individual transitions of \hit\ (black spectra) with a Voigt profile fit (red-blue solid lines). Note the large number  of fitted \hit\ transitions and the lack of any evidence for contamination in these transitions.  \label{f-hifit}}
\end{figure}

For the PF, we used a modified version of the software described in \citet{fitzpatrick97}, which is described in more detail in \citet{lehner11a} and \citet{lehner14}. The same continua used in the COG and AOD methods were also used for the PF (i.e., the continuum placement was not a free parameter in the PF). A major difference from the previous methods is that the model profiles were convolved with the COS instrumental line-spread function. The three parameters for each component $i$ (typically $i=1$, but in some cases $i=2$ or 3) --- column density ($N_i$), Doppler parameter ($b_i$), and central velocity ($v_i$) --- are input as initial guesses and were subsequently varied to minimize $\chi^2$ of the fit to all the fitted \hi\ transitions. In some rare cases the alignment of one or two transitions in a given absorber was somewhat off ($\ga 20$ \km\ relative to other transitions), and in these cases the relative velocity of the departing transition was also allowed to vary during the fit. As for the COG, we started with all the \hi\ transitions that did not appear {\it a priori} contaminated and iterated, removing any transitions that {\it a posteriori} appear contaminated or where a close absorption feature lies within the velocity range over which the $\chi^2$ is estimated (we note that we could mask some of these features, but in all the cases we had enough \hi\ transitions to model accurately \nhi\ that we did not undertake this extra and more time consuming step). In Fig.~\ref{f-hifit}, we show an example of profile fitting that shows no contamination in many \hi\ Lyman series transitions. Our adopted column densities from the PF are summarized in Table~\ref{t-results} in the FIT line entry.

In general, there was a good agreement between these different methods and we always use a combination of at least two of these methods and in many cases the 3 methods, to estimate \nhi\ for absorbers with  $15.2 \la \mlnhi \la 16.9$.  The three methods explore different parameters and different transitions, allowing us to reliably estimate  \nhi. We simply averaged the values and propagated the errors accordingly. Our adopted \nhi\ results are summarized in  Table~\ref{t-results} in the AVG line entry.

Finally, for absorbers with  $ \mlnhi \ga 18$, we generally could derive only a lower limit with the break method at the Lyman limit. In this case, we had to use further constraints to accurately determine \nhi, as the higher-order Lyman series lines remain saturated for many more transitions than the weaker \hi\ absorbers. Only for absorbers where we have simultaneously \lya\ and/or \lyb\ and very low \hi\ transitions could we derive relatively accurate \nhi\ in that column density range (typical errors on \nhi\ are of order 0.2--0.3 dex for these strong \hi\ absorbers). The \lya\ absorption profile is particularly critical because at these amounts of \hi\ the line profile begins to exhibit damping wings. The weaker transitions were used to make sure that the profile was not over-fit and the absence of some of the weakest \hi\ transitions helped also putting a lower limit on \nhi. Only 5 absorbers had the required transitions and signal-to-noise to accurately estimate \nhi\ at $ \mlnhi \ga 18$, and are summarized in Table~\ref{t-results}. There were 11 additional absorbers where we could derive only a lower limit from the break and one that was originally \mgii-selected (and hence is not in our final sample) that we list in Table~\ref{t-hifail} for completeness. 

As noted in \S\ref{s-search}, there are two recent surveys that were undertaken with a smaller sample of COS data where  \nhi\ was estimated \citep{stevans14,shull17}. The updated \nhi\ values in \citet{shull17} are in agreement with our own independent estimates within 1--$2\sigma$, except for a few cases where the results from \citet{stevans14} are in better agreement (e.g., J124511.26+335610.1 at $\mzabs = 0.556684$). We also note that some differences between these and our survey result from a different treatment of the velocity components whereby they sometimes combine more than 1 component in their estimate of the column densities (when we add the column densities of the various components in our survey, we would find essentially the same results). 

Although it is always possible to have some errors in a large survey, we examined and analyzed the absorbers in a systematic way and we used several methods to estimate \nhi, which explore different parameters and different transitions, adding confidence to our results. We also note that for the L13 sample that was revisited here, we found results that consistent within less than $1$--$2\sigma$ despite using different sets of data (i.e., different data reduction and co-addition of the individual exposures) and an independent analysis.  

\section{Results}\label{s-results}
In this paper, we present results stemming directly from the estimated properties of the absorption for each absorber. In the subsequent papers, we will use photoionization models to interpret the data and, in particular, to estimate the metallicities of the absorbers. With just the column densities of metals and of \hi, we show below that the gas with $15.1 \la \mlnhi < 19$ is predominantly ionized; that there is evidence of multiphase gas; and that this gas has only very mild depletions of the heavy element Fe, to levels less than is observed in the Milky Way halo.

\begin{figure}[tbp]
\epsscale{1.2}
\plotone{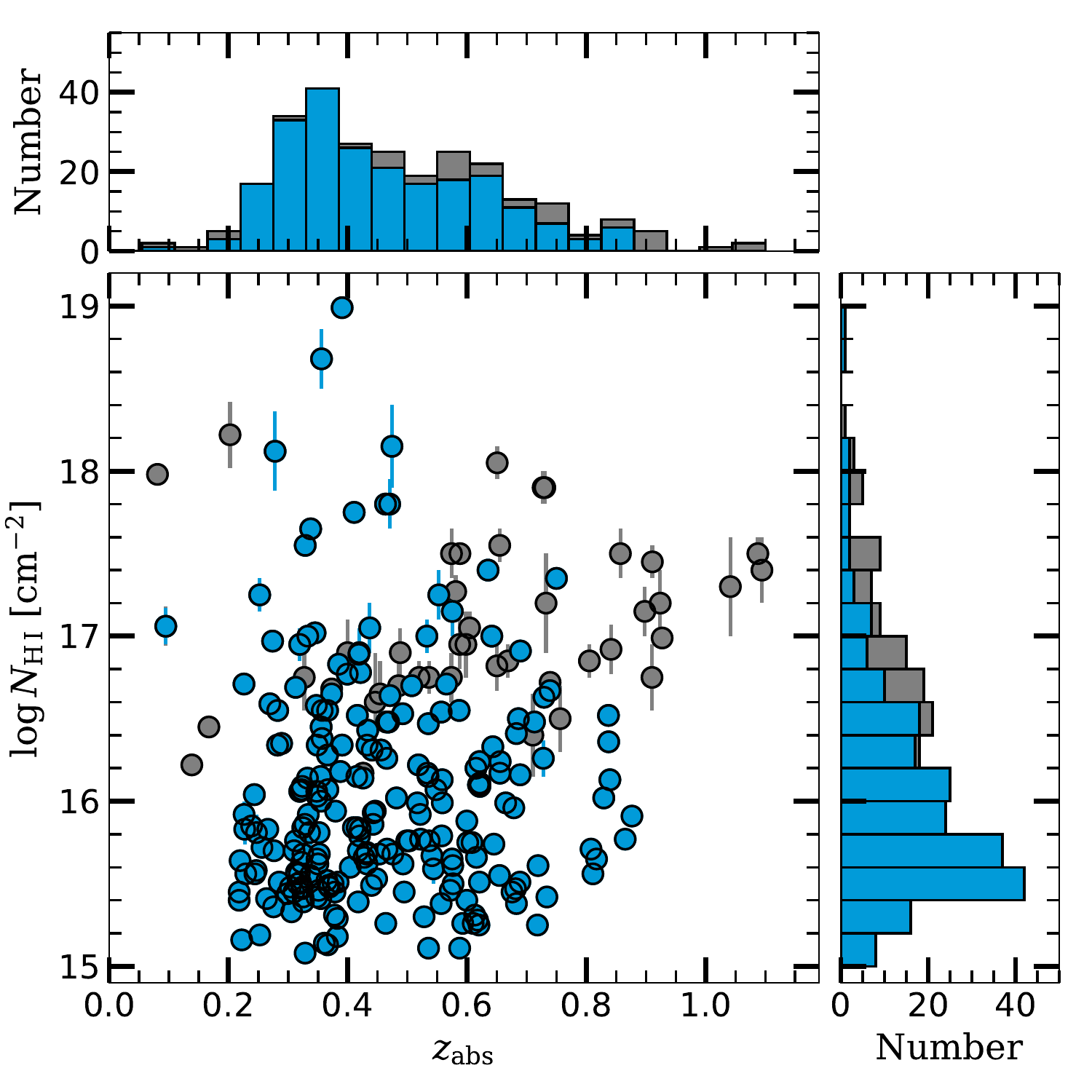}
    \caption{Scatter plot showing the distribution of the redshifts and \hit\ column densities for the absorbers in our sample (blue data). The gray data show additional data from the  samples analyzed in \citetalias{lehner13} and  \citetalias{wotta16} that we did not revisit in this work. The top and right panels show the redshift and $\mlnhi$ distributions with the same color-coding definition. 
    \label{f-nhi-vs-z}}
\end{figure}

\subsection{\hi\ Column Density and Redshift Distributions of the Absorbers}\label{s-res-nhi-z-dist}
In Fig.~\ref{f-nhi-vs-z}, we show the distributions of the absorber redshifts and \hi\ column densities. For the sample analyzed in this paper (shown in blue in this figure), the minimum ($z_{\rm min} \sim 0.2$) and maximum ($z_{\rm max} \sim 0.9$) redshifts  are set by the observational constraints (COS wavelength bandpass and coverage of the \hi\ Lyman series).\footnote{The only exception is the absorber at $\mzabs = 0.09487$ observed toward J130532.99-103319.0 for which the \nhi\ estimate comes from \fuse\ observations (see \citealt{cooksey08}), but we analyzed here the metal transitions that are in the COS bandpass. \label{f-except}} The redshifts largely overlap between the previous sample (i.e., the sample analyzed only in \citetalias{lehner13} and  \citetalias{wotta16}) and the newly analyzed sample here, except for a handful of absorbers at $0.9 \la \mzabs \la 1.1$, which mostly come from the \citetalias{wotta16} sample. The mean redshift and standard deviation of the entire sample (blue and gray data points in Fig.~\ref{f-nhi-vs-z}) are $\langle \mzabs \rangle = 0.48 \pm 0.19$ (the median being $\mzabs = 0.44$).  Although the redshift distribution slightly peaks around $0.3\la \mzabs < 0.4$, the overall redshift distribution is quite uniform in the range $0.2 \la \mzabs \la 0.7$. There is also no trend between $\mzabs$ and \nhi, i.e., a nearly 3 dex dispersion in \nhi\ is observed at any redshift in the interval $0.2 \la \mzabs \la 0.9$, although we note the lack of strong LLSs with $\mlnhi \ga 18$ at $\mzabs \ga 0.5$ (see below). 

For \nhi, the entire sample ranges from $10^{15.08}$ to $10^{18.99}$ \cmm. Absorbers with $\mlnhi \la 16.2$ are only found in the new sample as our previous surveys did not target these absorbers. To the best of our knowledge, none of the absorbers with  $15.1 \la \mlnhi \la 17$ were specifically targeted. These absorbers also do not produce a major break in the flux of the QSO spectra or produce major \hi\ absorption features in low-resolution spectra and hence the background QSOs were potential targets to be observed with COS G130M and/or G160M.  Therefore, the distribution of these absorbers should follow the overall distribution of the \hi\ absorbers in column density, which can be described as a power law $f_{\rm N_{HI}} \propto N_{\rm H\,I}^{-\beta}$ at $z<1$ (e.g., \citealt{lehner07,danforth16}); for a subset of the sample presented here, \citet{shull17} derived $\beta = 1.48 \pm 0.05$ for absorbers with $15 \la \mlnhi \la 17$. The number of absorbers at $\mlnhi \la 15.3$--15.4 drops sharply; this is a pure artifact from our survey design since we are not complete below this threshold owing to the lack of sensitivity in the COS data to estimate low metallicities. 

For absorbers with $17 \la \mlnhi \la 19$ and observed with COS G130M/G160M observations, the situation is more complicated since these absorbers can absorb part of or the entire UV flux, and many QSOs with known strong LLSs observed in low-resolution UV spectra might not have been targeted with the high-resolution mode of COS. However, our blind search of these LLSs identified more systems than we initially expected at least at $\mzabs \la 0.5$. We attribute this to the fact that many QSOs with no existing UV spectra prior to the COS observations were most certainly selected based on the FUV and NUV magnitudes obtained from the {\it Galaxy Evolution Explorer} ({\it GALEX}) mission. Since the {\it GALEX}\ FUV magnitude peaks around 1525 \AA\ (and is most sensitive around 1420--1650 \AA), a strong LLS can be present at $z\la 0.5$ without any way to know prior to the acquisition of a COS spectrum. And, indeed, all the absorbers with $\mlnhi \ge 17.5$ (i.e., optically thick at the Lyman limit, $\tau_{\rm LL}\ge 2$) in the new sample of strong LLSs have $\mzabs < 0.5$ (although we note that 4 strong \hi\ absorbers for which we could derive a lower on \nhi\ were found at $z>0.5$, see Table~\ref{t-hifail}; these may have been selected using a NUV or optical selection). \citet{shull17} found the same effect in their smaller sample where they did not find any LLSs with $\mlnhi \ge 17.2$ at $\mzabs > 0.45$, attributing also this bias to the {\it GALEX}\ FUV pre-selection of the targets. We note, however, that the survey undertaken by \citetalias{wotta16} originally aimed to study the frequency of the LLSs at $0.4 \le \mzabs \le 1$ selected the background QSOs based on their visual magnitude, and hence did not suffer from selection biases.  

Therefore, our survey in the $15.4 \la \mlnhi \le 17.2$ range should follow the column density distribution of \hi\ absorbers. In the range $17.2 < \mlnhi < 19$, there is a bias against these absorbers at $0.5\la \mzabs \la 0.9$, but not at lower redshift. However, in this redshift interval, we have also several absorbers from our COS G140L survey \citepalias{wotta16} (this survey covers only absorbers in the $0.4 \le \mzabs \le 1$) where there is no bias against strong LLSs. Hence, the combined sample of absorbers with $15.4 \la \mlnhi \la 18$  should also follow closely the column density distribution at $z\la 1$.

\begin{figure*}[tbp]
\plottwo{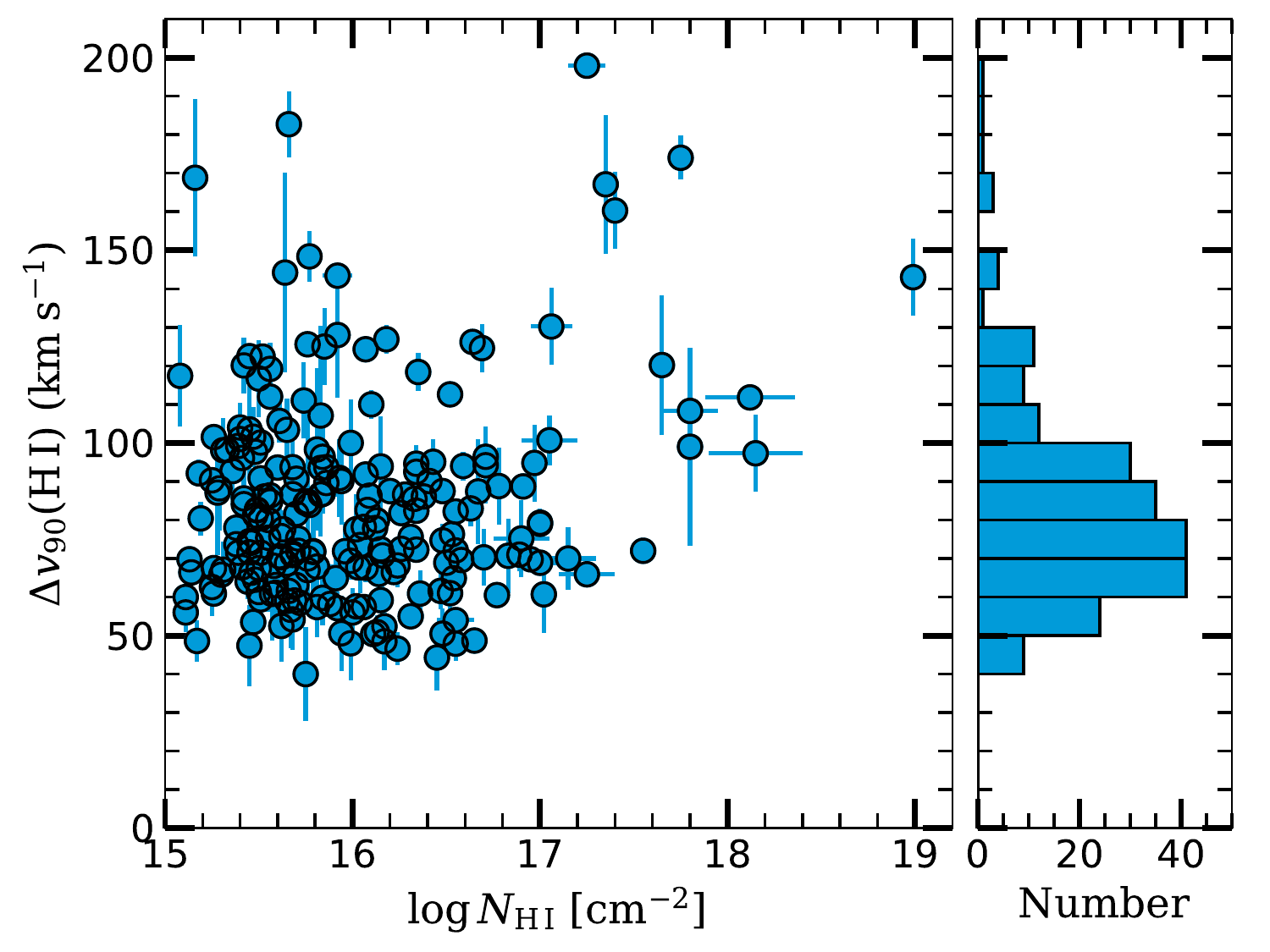}{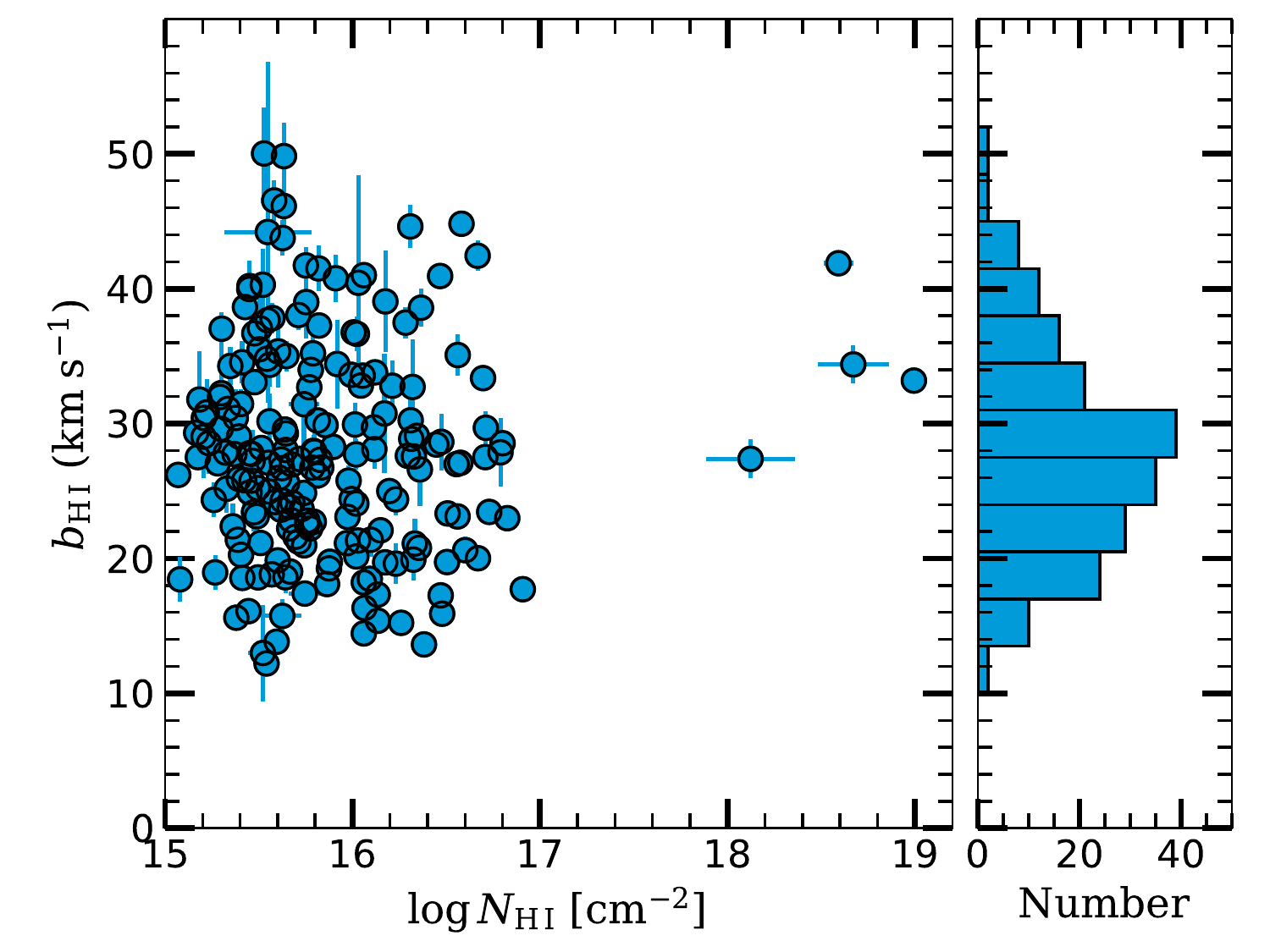}
    \caption{{it Left}: Comparison of the velocity width against the \hit\ column density (\dv\ is obtained from the integration the velocity profiles). {\it Right}: Comparison of the Doppler parameters and \nhi\ for the absorbers that were profile fitted. 
    \label{f-dv90-hi-z}}
\end{figure*}

\subsection{Velocity Widths and Doppler Parameters}\label{s-res-vel}
We now consider the velocity widths of \hi, low ions (\cii\ and \mgii), and intermediate ions (\ciii). The velocity widths, \dv, were estimated using the same integration ranges that were used to determine the column densities. Although we provide information about the \ovi\ (and other high ions), we remind the reader to treat these results with caution as we only integrated the profiles over the same velocity range. As shown by \citet{fox13} for the L13 sample of pLLSs and LLSs, the \ovi\ profiles are typically broader than the low-ion profiles and their velocity centroids are offset from \hi, indicating that \ovi\ and \hi\ do not necessarily coexist in the same regions (or have different weightings with velocity). This conclusion appears supported by the larger sample presented after inspecting the absorption velocity profiles in the Appendix, but further quantifying the differences between high and low ions is beyond the scope of our survey (but will be part of paper IV).

\begin{figure*}[tbp]
\epsscale{0.9}
\plotone{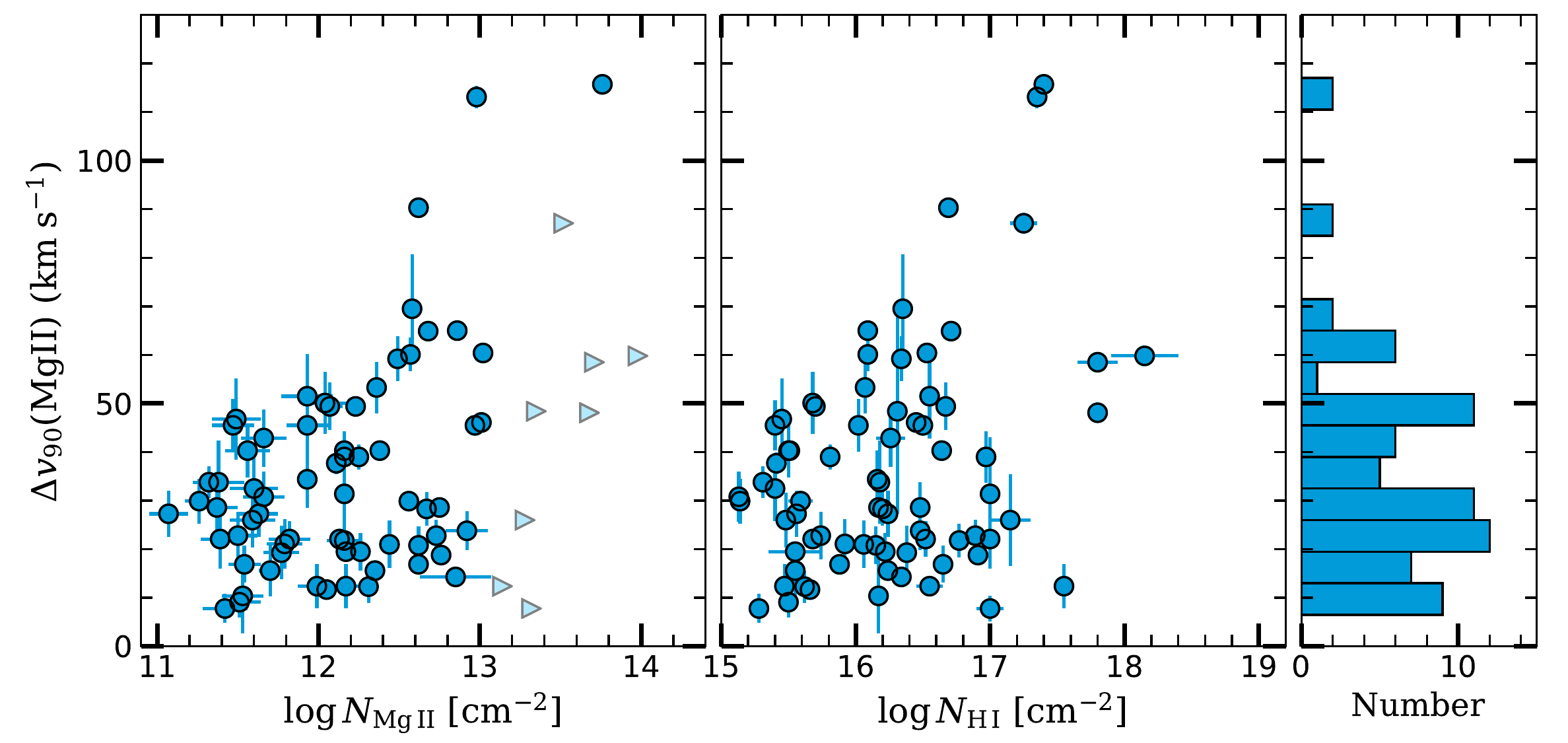}
    \caption{Comparison of the velocity width of \mgiit\ against the \mgiit\ and \hit\ column densities.  
    \label{f-mgiidv90}}
\end{figure*}

In the left panel of Fig.~\ref{f-dv90-hi-z}, we compare the velocity width (\dv) of \hi\ with \nhi. The velocity width ranges from 40 to 198 \km, with 86\% of the absorbers having $50 \le \mdv \le 120 $ \km\ and the mean (and standard deviation) being $84 \pm 27$ \km. The lower bound is unlikely to be an artifact since 40--50 \km\ is 2--3 times the COS G130M and G160M resolution and since lower $\mdv$ are  found in metal lines.  For absorbers with $\mlnhi \la 17$, there does not appear to be any trend between the velocity width and \nhi, and only 15\% of the absorbers have $\mdv > 100$ \km\ in that \nhi\ range.  In contrast,  65\% of the absorbers with $17 \la \mlnhi \la 19$ have  $\mdv > 100$ \km. This reflects in part  our inability to separate  components at larger \nhi. However, there is no absorber in our sample with $\mdv \ga 200$ \km, implying that these are extremely rare at $z\la 1$ (but not at higher $z$, see \citealt{lehner14,lehner17} and \S\ref{s-prox}).  

In the right panel of  Fig.~\ref{f-dv90-hi-z}, we compare the Doppler parameter, $b$, and \nhi\ derived from the profile fitting in the individual components of \hi. As for \dv, there is no trend between $b$ and \nhi, which differs from the trend seen at $\mlnhi < 15$, where an increase of $b$ with \nhi\ is observed (although with a larger scatter, see \citealt{lehner07}).  The  mean and median values are $28$ \km\ and $27$ \km, respectively, while the standard deviation is $7.8$ \km. The mean and median values are smaller than those seen in the \lya\ forest (with $13.2 \le \mlnhi \le 14$) at similar redshifts where about 30\% the components consist of broad absorption ($b>40$ \km) \citep{lehner07}.  In contrast, in our sample, 91\% of the profile-fitted absorbers have $b<40$ \km\ and 16\% have even $b<20$ \km. We, however, note that the $b$-values for the 8 absorbers with $\mlnhi\ga 15$ in the \citet{lehner07} high-resolution survey ranges from about 15 to 50 \km, fully consistent with our much larger survey. 

The Doppler parameter for \hi\ is related to the gas temperature via  $b_{\rm H\,I} = (2kT/m_{\rm H} + b^2_{\rm nt})^{0.5}  $, where $k$ is the Boltzmann constant, $T$ the temperature of the gas, $m_{\rm H}$ is the mass of hydrogen, and $ b_{\rm nt}$ is the (unknown) non-thermal component to the broadening. Hence for $\langle b \rangle = 28$ \km, the temperature of the gas is $T<5 \times 10^4$ K and consistent with the gas being primarily photoionized. 

\begin{figure*}[tbp]
\epsscale{1.2}
\plotone{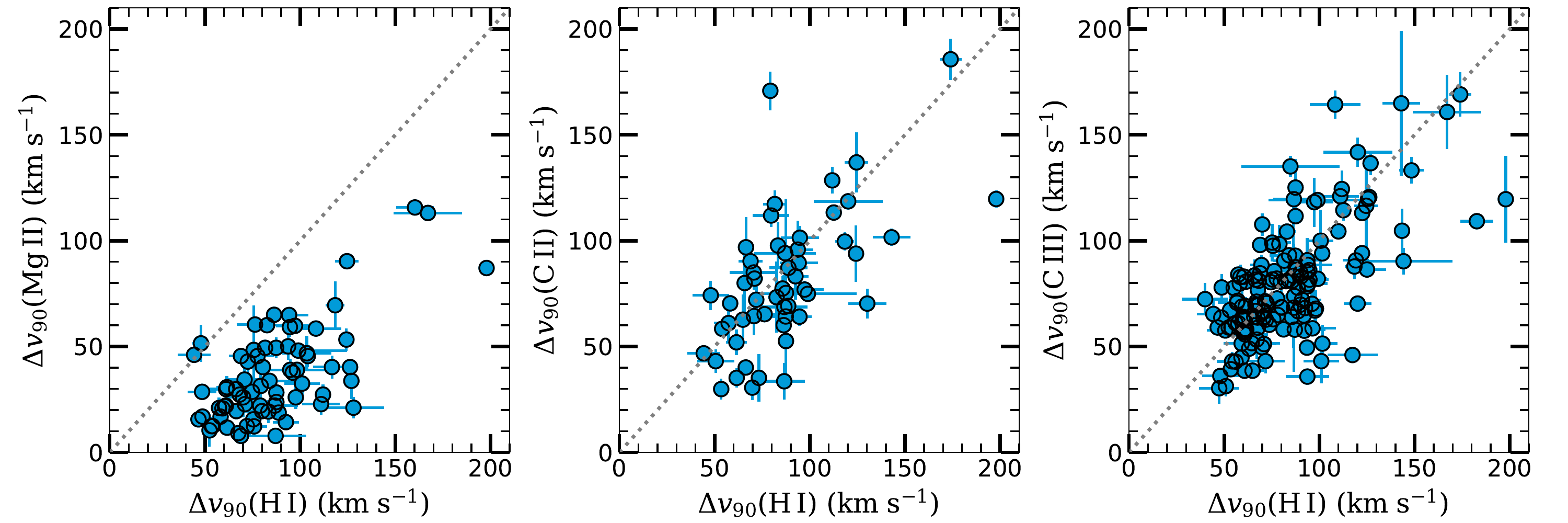}
    \caption{Comparison of the velocity widths of \mgiit, \ciit, and \ciiit\ against that of \hit. The dotted line in each panel is the 1:1 relationship. Observations of \mgiit\  have a resolution a factor $\sim$2.2 better than the COS observations (\ciit, \ciiit, \hit), explaining the observed difference.   \label{f-dv-vs-dv}}
\end{figure*}

In Fig.~\ref{f-mgiidv90}, we show the velocity width of \mgii\ against the column densities of \mgii\ and \hi. Seventy six per cent of the absorbers have \dv\ values for \mgii\ in the 10--50 \km\ range, a factor $\sim$2 smaller than that of \hi, but, as we argue below, this is mostly an instrumental resolution effect. Using the Spearman rank-order, the test confirms the visual impression that there is a moderate positive monotonic correlation between $\mdv({\mbox \mgii})$ and the column densities of \mgii\ and \hi\ (Spearman correlation coefficient $r_{\rm S} = 0.43$ and 0.36 with a $p$-value\,$< 0.1\%$ respectively).\footnote{We adopt the following guide for the absolute value of  $r_{\rm S}$: 0.00--0.19: very weak, 0.20--0.39: weak, 0.40--0.59: moderate, 0.59--0.79: strong, and 0.80--1.00: very strong monotonic relationship.} However, there is not a general increase of $\mdv({\rm Mg\,II})$ with \nmgii: large velocity profiles ($b>50$ \km) are only observed when $\mlnmgii\ \ga 12.5$, i.e., as \nmgii\ becomes larger, the number of \mgii\ components often increases. However, even as the column density of \mgii\ increases, a large fraction of the absorbers has also narrow and simple absorption profiles. The same trend is observed when comparing  $\mdv({\mbox \mgii})$  and \nhi.

Finally in Fig.~\ref{f-dv-vs-dv}, we compare the velocity widths of  \mgii, \cii, and \ciii\ with \hi. There is a strong correlation between \dv\ of \mgii, \cii, and \ciii\ with \hi\ (Spearman correlation test $r_{\rm S} \simeq 0.65$ with $p \ll 0.1\%$). For \cii\ and \ciii\ with \hi, the data are scattered around the 1:1 relationship. For \mgii, the data are scattered below the 1:1 relationship. In Table~\ref{t-dv90}, we give the mean, standard deviation, and median of \dv\ for all these species and \siiii. On average, \dv\ for \mgii\ is a factor $\sim$2.2 smaller than that of \cii, \ciii, or \hi\ (\dv\ means, medians, and scatter are effectively the same for \hi, \cii, and \ciii), which is essentially the difference between the COS and high-resolution ground-based observations. Hence the difference observed between \mgii\ and UV metal-line UV transition is an instrumental artifact, but it also implies that the typical values of \dv\  for the metal lines in the absorbers with $15< \mlnhi < 19$ is close  to $36 \pm 22$ \km.\footnote{We also note that \dv\ for \hit, \siiiit, \ciit, and \ciiit\ are essentially the same on average, implying that the atomic mass of species does not play a significant role in the broadening.}

Although we do not show the comparison between the velocity centroids of \hi\ and of \cii, \ciii, and \mgii\ (and other low and intermediate ions), the dispersion in the centroid offsets between \hi\ and low/intermediate ions is scattered around 0 \km\ (see also \citealt{fox13}). All these arguments confirm the visual inspection of the absorption profiles that the \hi\ and the low/intermediate trace the same gas and that the gas temperature is cool with $T\la 5 \times 10^4$ K. 
 
\subsection{Column densities}\label{s-res-col}
\begin{figure}[tbp]
\epsscale{1.15}
\plotone{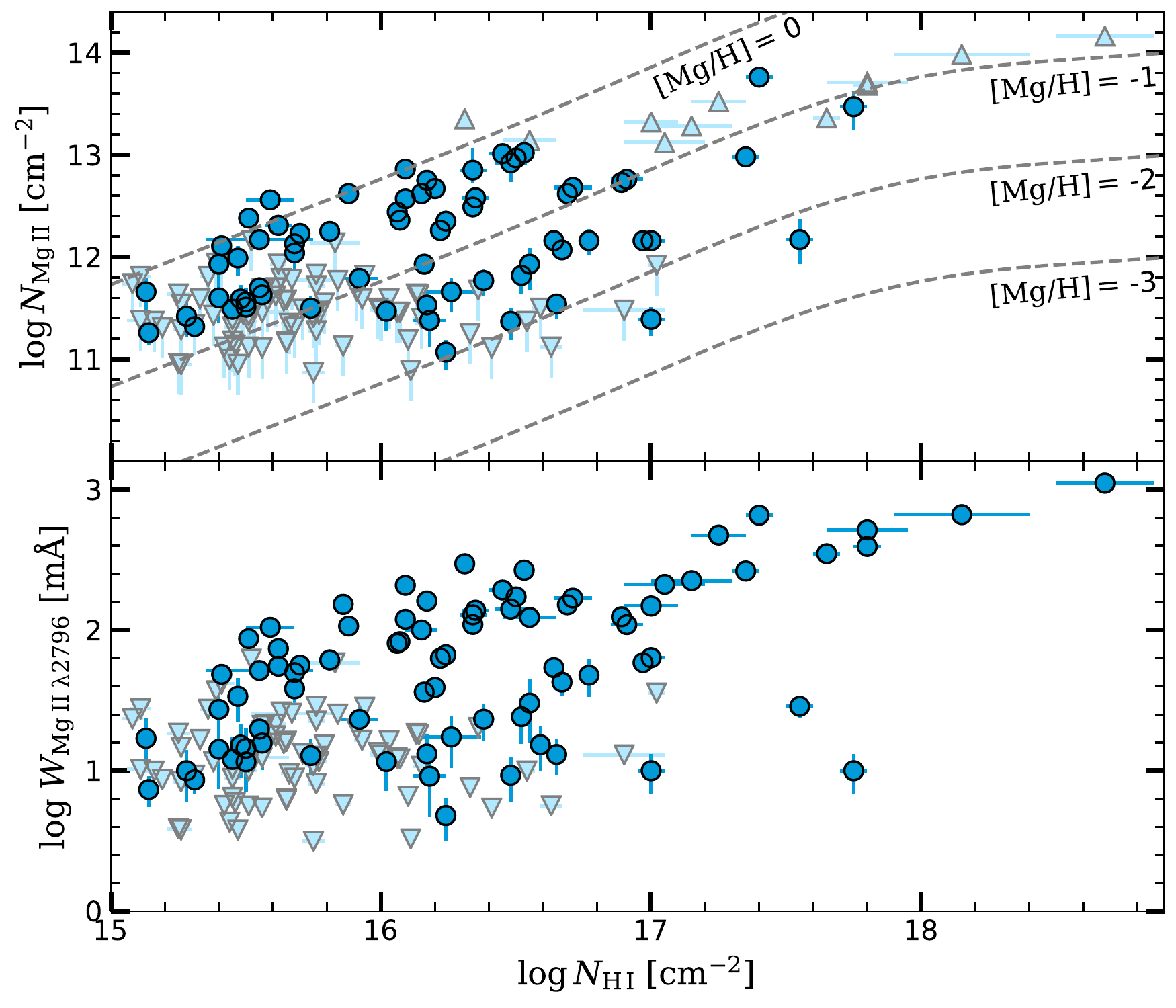}
    \caption{Column density of \mgiit\ ({\it top}) and equivalent width of \mgiit\ $\lambda$2796 ({\it bottom}) against the column density of \hit. Down triangles are $2\sigma $upper limits, while triangles are lower limits. In the top panel, the dashed lines represent the behavior of \nmgii\ with \nhi\ in a photoionized gas with a solar ({\it top}) to $10^{-3}$ ({\it bottom}) solar metallicity. 
\label{f-mgiivsnhi}}
\end{figure}

With our results in hand, one of the most straightforward things to ask is how the metal ion column densities relate to those of \hi. In Fig.~\ref{f-mgiivsnhi}, we compare the column densities and equivalent widths of \mgii\ with the column densities of \hi. The ranges of \nmgii\ and $W_{\rm Mg\,II}$ are between $<$11 and $>$14 dex and $<$3--1100 m\AA, respectively. There is a visual strong correlation for both \nmgii\ and $W_{\rm Mg\,II}$ with \nhi, which is confirmed by the Spearman correlation test $r_{\rm S} \simeq 0.71$ and 0.57, respectively, ($p \ll 0.1\%$; samples without upper limits --- if the entire sample is considered instead, then $r_{\rm S} \simeq 0.61$). The values at $\mlnmgii \la 12$ ($W_{\rm Mg\,II} \la 40$ m\AA) are nearing our detection limit. For $\mlnmgii \ga 12$ ($W_{\rm Mg\,II} \ga 40$ m\AA), there is a strong positive correlation between \nmgii\ and $W_{\rm Mg\,II}$ with \nhi. There is a large range of \nmgii\ at all \hi\ column densities. For example, over the range $16.0 \la \mlnhi \la 16.5$ there is a $>$2 dex spread in the observed values of \nmgii.

\begin{figure*}[tbp]
\epsscale{1}
\plotone{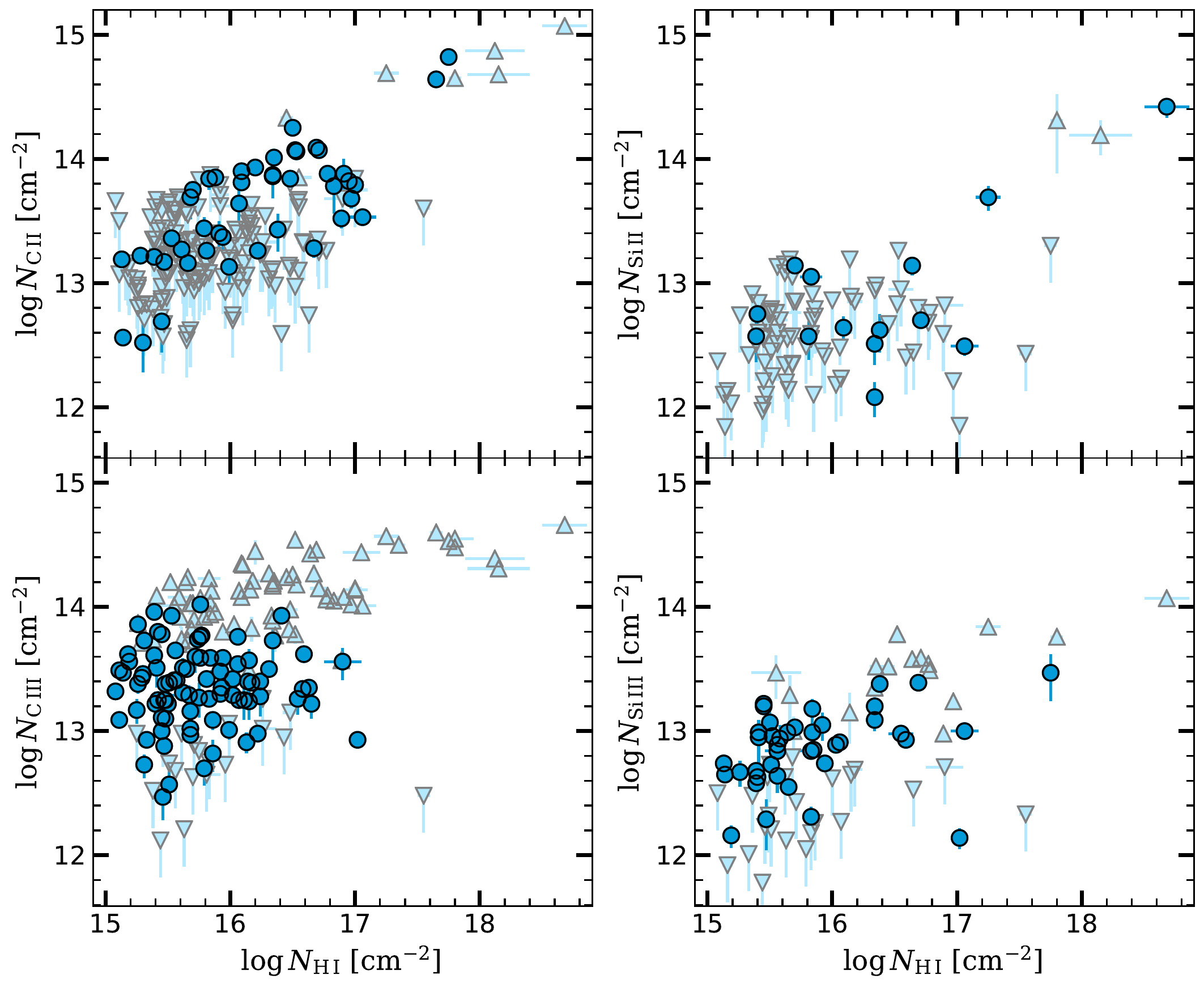}
    \caption{Column densities of selected key metal ions against the column density of \hit. Down triangles are $2\sigma$ upper limits, while triangles are lower limits. \label{f-ionsvsnhi}}
\end{figure*}

In  Fig.~\ref{f-mgiivsnhi}, we also show as dashed lines in the top panel the general trends of metallicities from solar to $-3$ dex solar, which are adopted from the photoionization models shown in Fig.~5 of \citetalias{wotta16}. We have not updated these models and only use them here as guides to better understand the origin of the observed trends between \mgii\ and \hi. As discussed in \citetalias{lehner13} and \citetalias{wotta16}, the strength of \mgii\ depends strongly on \nhi\ and the metallicity, but only weakly on the strength of the ionizing background. This is because the ionization correction for \mgii\ does not vary strongly over the range of ionization parameters probed by these absorbers owing to the similarity of the ionization potentials of \hi\ and \mgii\ and a lack of strong spectral features in the UV radiation fields \citepalias{wotta16,lehner13}. As such, \mgii\ is a particularly good indicator of metallicity in the studied \nhi\ range. The overall increase of \nmgii\ with increasing \nhi\ is seen for absorbers with $-1 \la \xh \la 0$, i.e., metal enriched systems. When $\mlnhi \ga 17$, the \mgii\ $\lambda \lambda$2796, 2803 doublet becomes more easily saturated, and the data cluster around the $\xh \simeq -1$ line (as we will see in paper II, with the addition of other ions we can more accurately and robustly estimate the metallicities for these absorbers). The scatter observed at $\mlnmgii \la 12$ is in part due to a sensitivity limit to low metallicity absorbers with \mgii, as there is a much larger fraction of upper limits when $15.2 \la \mlnhi\ \la 15.7$; in that range we can still detect metal-poor absorbers with $\xh \la -1$, but not much below that (again this applies to \mgii\ alone; as we show in paper III, we can push below this limit using other metal ions). The lack of data around the $\xh \simeq -1$ line for absorbers with $16 \la \mlnhi \la 17$ is particularly interesting as it is around this value that the dip in the bimodal metallicity distribution of the pLLSs is observed (\citetalias{lehner13, wotta16}; paper II). Interestingly, this feature is also observed clearly for the equivalent widths of \mgii\ where a hole in the scatter is seen for absorbers with $16 \la \mlnhi \la 17$. This demonstrates that  the bimodal metallicity distribution in the $16 \la \mlnhi \la 17$ range is not artifact of the ionization modeling. For $ \mlnhi < 15.9$, we would need much better SNR ground-based spectra to improve the current limits on \mgii.

In Fig.~\ref{f-ionsvsnhi}, we compare the column densities of \cii, \siii, \ciii, and \siiii\ with \nhi. Without the upper limits, the Spearman correlation test gives $r_{\rm S} \simeq 0.83$ and 0.75 ($p \ll 0.1\%$) for \cii\ and \siii, respectively, implying a very strong correlation between the column densities of the singly ionized species with \nhi\ (these ranking numbers drop to 0.64 and 0.54, respectively, if upper limits are included). The sample with detections of \cii\ is much smaller than that of \mgii\ and the upper limits are not as stringent, but there is also some evidence for a gap in the \ncii\ distribution in the $16 \la \mlnhi \la 16.9$  range. For \ciii\ and \siiii, the correlation is not as strong, but still significant with $r_{\rm S} \simeq 0.5$ ($p \ll 0.1\%$). While there is a correlation between all these ions and \hi, at any given \nhi\  in the range $15 \la \mlnhi \la 17$, there is a factor 10--30 (1--1.5 dex or more since some column densities are lower limits) scatter in the column densities of the metal ions. At higher \nhi\ ($\mlnhi \ga 17.5$), except for one absorber, high metal column densities are observed. Both ionization and metallicity affect these column densities (and in particular \ciii\ and \siiii\ are more dependent on the ionization parameter than the singly ionized species); however, as shown by \citetalias{lehner13} and \citetalias{wotta16} (and see papers II and III), the ionization parameter $U$ does not change drastically in the \hi\ column density range $15 \la \mlnhi \la 19$. As implied by the analysis of \citetalias{wotta16} and confirmed with a larger sample in paper II, very low metallicity absorbers ($\xh \la -1.4$) are rarer at $\mlnhi \ga 17.5$, while common at lower \nhi, which explains in part the low and high scatter in the metal lines above and below $\mlnhi \simeq 17.5$,  respectively. 

\begin{figure*}[tbp]
\epsscale{1}
\plotone{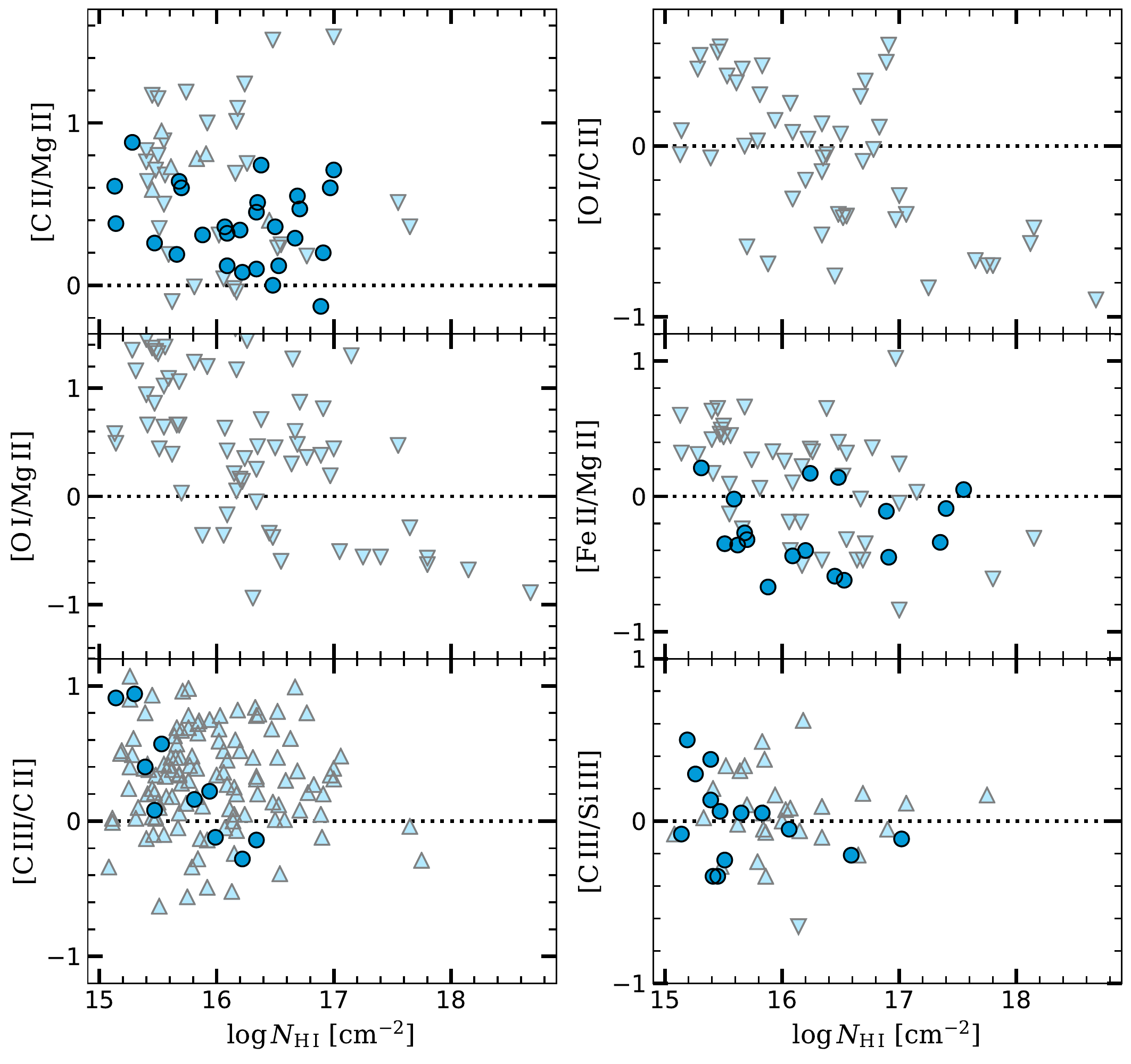}
    \caption{Ratios of various ions and atoms against the column density of \hit\ (error bars are not plotted but are, on average, about 0.10--0.15 dex on the $y$-scale). Down triangles are $2\sigma$ upper limits, while triangles are lower limits. The dotted line indicates the solar relative abundance in each panel. This set of panels implies that the gas is predominantly ionized and dust depletion is negligible for the absorbers in our sample. 
 \label{f-ratiosvsnhi}}
\end{figure*}

Figs.~\ref{f-mgiivsnhi} and \ref{f-ionsvsnhi} also give a measure of the sensitivity of our survey for various key ions. For the singly ionized species, \mgii\ (with many upper limits at the level or lower than the detections) fairs typically better than \cii\ and \siii\ (where many upper limits are quite high and overlap with relatively high columns of \cii\ or \siii). The major difference is that the ground-based observations have a more uniform SNR than the COS observations. For \ciii, its $\lambda$977 transition is so strong that the upper limits are scattered among the lowest column densities where \ciii\ absorption is detected. \ciii\ is also the ion with the lowest fraction of non-detections (11\% of the sample) compared to 73\%, 82\%, 50\%, and 30\% for \cii, \siii, \mgii, and \siiii, respectively. 

Typically, when \ciii\ is not detected, no other metals are detected. The exception is the absorber toward J135726.26+043541.3 at $z = 0.328637$ ($\mlnhi = 17.55\pm 0.05$), for which there are weak detections of the strong NUV transitions of \feii\ and \mgii. As we will see in papers II and III, absorbers with no \ciii\ are among the lowest metallicity absorbers. Looking at Fig.~\ref{f-ionsvsnhi}, there are only 2 absorbers with no \ciii\ (and no other metal ion detections) where the upper limits are well below the lowest detections. This implies that truly pristine gas at $z\la 1$ is extremely rare with $<$3\% (90\% confidence level) of the absorbers with no metal detection. We will show in papers II and III that there is in fact no evidence for gas with $<$1/1000 solar metallicity at $z\la 1$. At $z\sim 2$--3.5, pristine gas is rare too but not inexistent with 3--18\% of the pLLSs/LLSs having no metals detected at levels $<1/10,000$ \citep{lehner16,fumagalli11b}. As shown in \citetalias{lehner13} and \citetalias{wotta16} and see also Fig.~\ref{f-mgiivsnhi},  there is, however, a significant fraction of the absorbers that have metallicities $<$1/100 solar at $z\la 1$, implying little enrichment for these absorbers over several billion years when comparing to the metallicities of the same absorbers at $z\sim 2$--3.5 \citep[see,][]{lehner16}.

\subsection{Column density ratios}\label{s-res-ratios}
In Fig.~\ref{f-ratiosvsnhi}, we plot several ionic/atomic ratios as a function of \nhi. Depending on the species compared, these ratios inform us about the ionization or depletion levels of the gas. As we discuss below, some of these elements may also be affected by the nucleosynthetic history of the gas. We use the usual squared-bracket notation, $[{\rm X}^{+i}/{\rm Y}^{+j}] \equiv \log (N_{{\rm X}^{+i}}/N_{{\rm Y}^{+j}}) - [{\rm X/Y}]_\sun$. Here X$^{+i}$ and Y$^{+j}$ represent elements X and Y in independent ionization stages $i$ and $j$, while $[{\rm X/Y}]_\sun$ represents the relative solar abundances of these elements (we use for the solar abundances from \citealt{asplund09}). No ionization corrections are applied in making these plots, so these values are sensitive both to relative ionization and abundance variations. Plotting these ratios as a function of \nhi\ allows us to determine if there is any trend with \nhi\ that could affect the metallicity comparison from low to high \nhi\ values. 

For the \hi\ column densities probed in our sample, [\cii/\mgii] is unlikely to be affected by dust depletion. In the Milky Way halo or the diffuse environment of the Magellanic Bridge, neither C nor Mg is strongly depleted  and their relative ratio should be roughly solar \citep[e.g.,][]{welty99a,lehner01a,lehner02,lehner08,jenkins09}. As discussed in \citetalias{lehner13} (and see also \citealt{lehner16} for a similar behavior at high redshift), the $[{\rm C}/\alpha]$ ratio is sensitive to nucleosynthesis effects with a time lag between the production of $\alpha$ elements (e.g., O, Si, Mg) and carbon. However, as shown in \citetalias{lehner13} and further demonstrated with a larger sample in papers II and III,  $[{\rm C}/\alpha]$ is on average around the solar value with some scatter (about 0.35 dex). On average, results from Fig.~\ref{f-ratiosvsnhi} show that  [\cii/\mgii$] = +0.37$. The super-solar value for  [\cii/\mgii] is consistent with the gas being largely ionized since the ionization energies (IE) of \cii\ are in the range 11.3--24.4 eV compared to \mgii\ 7.7--15.0 eV. 

Similarly to C and Mg, O is very mildly depleted onto dust and the relative abundance of O relative to C or Mg should be about solar. Fig.~\ref{f-ratiosvsnhi} shows that \oi\ is not detected in the \hi\ column density range $15 \la \mlnhi \la 19$, which is already a signature by itself that the gas must be largely ionized. Furthermore, for several absorbers, we were able to place strong upper limits on \oi\ so that [\oi/\mgii] or [\oi/\cii$]\ll 0$, demonstrating that the gas is predominantly ionized without any ionization modeling \citep{lehner01b,lehner01a}.  For most of the absorbers, Fig.~\ref{f-ratiosvsnhi} shows that [\ciii/\cii$]\ga 0$, also directly implying that the gas is ionized. 

[\ciii/\siiii] is scattered around the solar value, which is consistent with the gas being photoionized by an ionizing source that is not dominated by a hard ionizing spectrum (e.g., dominated by QSOs), otherwise we would expect [\ciii/\siiii$]\gg 0$ (the IE range for \ciii\ is 24.4--47.9 eV compared to 16.3--33.5 eV for \siiii). The scatter in the  [\ciii/\siiii] ratio is consistent with the scatter we find in the [C/$\alpha$] ratio (see papers II and III). 

While ionization can affect the [\feii/\mgii] ratio, the effect should be small since they have similar ionization corrections over the densities probed by these absorbers and IE ranges  (IE 7.9--16.2 for \feii\ compared to \mgii\ 7.7--15.0 eV). The  [\feii/\mgii] ratio can be, however, affected by dust depletion, where Fe is typically more depleted onto dust than Mg \citep[e.g.,][]{savage96,welty99b,jenkins09}, or by nucleosynthesis, whereby an $\alpha$-element can be enhanced relative to Fe in metal-poor environments such as in DLAs \citep[e.g.,][]{rafelski12,quiret16}.  Only including the detections of \feii, we find $\langle [$\feii/\mgii$]\rangle = -0.25 \pm 0.27$. Using a survival analysis where the censored data (upper limits) are included \citep{feigelson85,isobe86}, we find  $\langle [$\feii/\mgii$]\rangle = -0.43 \pm 0.06$ (where the error is the error on the mean from the Kaplan-Meier estimator; the scatter being  $0.30$ dex). The scatter and mean are quite similar to the Fe/$\alpha$ seen in DLAs with $[\alpha/{\rm H}] \le -1.5$ where $\langle [{\rm Fe}/\alpha]\rangle = -0.5 \pm 0.3$ (see paper II).  At the metallicities probed by these absorbers, both nucleosynthesis and depletion effects may play a role. 

As mentioned above, iron is known to be depleted onto dust in the Milky Way, and it is the element that is the most affected by dust depletion in our sample  \citep{savage96,welty99b,jenkins09}.\footnote{The comparison of the relative interstellar abundances of Fe relative to weakly depleted $\alpha$ elements  in the Magellanic Cloud region \citep{welty97,welty99a,lehner01a,jenkins17} and those in the Milky Way \citep{jenkins09} reveals comparable depletion patterns, despite global differences in the metal and dust content of these environments, which implies we can use the Milky Way results as a guideline.} \citet{jenkins09} found in the Milky Way for the lowest depletion factor ($F_* = 0$), the observed depletion of Mg and Fe are about $-0.3$ and $-1$ dex, respectively, corresponding to [Fe/Mg$] \simeq -0.7$.\footnote{As the depletion factor increases,  [Fe/Mg] decreases (e.g., if $F_* = 0.5$, then [Fe/Mg$] = -1$).} Fig.~\ref{f-ratiosvsnhi} also shows no trend of  [\feii/\mgii] with increasing \nhi, which would be expected if the depletion was important.  The mean value  $\langle [$\feii/\mgii$]\rangle = -0.4 \pm 0.3 $ implies the Fe depletion in the gas probed by the absorbers with $15 \la \mlnhi \la 19$ at $z<1$ are lower than the least-depleted environments in the Milky Way and Magellanic Clouds. Since $\alpha$-enhancement can also affect this [\feii/\mgii] ratio, the dust depletion effect could be even smaller. Therefore since Fe is the most refractory species studied in our sample, it also implies that dust depletion is negligible for all the absorbers in our sample.  

Hence, there is some suggestion based on both [\feii/\mgii] and [\ciii/\siiii] (i.e., [C$/\alpha$]) that there is some nucleosynthetic enrichment history of the gas, in particular $\alpha$-enhancement characteristic of nucleosynthesis from Type II supernovae. However, this effect is mild and not observed systematically in all the absorbers. 
 
\subsection{Proximate Absorbers and Multiple Component Absorbers}\label{s-prox}
In our search for \hi-selected absorbers, we did not reject {\it a priori}\ absorbers from our sample if their redshift was near the redshift of the QSO, i.e., $\Delta v \equiv (z_{\rm em} - z_{\rm abs})/(1+z_{\rm abs})\, c < 3000$ \km, where $c$ is the speed of light. These are known as proximate absorbers (also known as ``associated" absorbers although this is a misnomer since these absorbers may or may not be actually associated with the QSO). If several absorbers are closely separated in redshift space (that we defined as paired absorbers, keeping in mind that more than two absorbers can be closely separated in redshift space), we considered them separately as much as possible. We identify these paired absorbers because, depending on our finding for their metallicities, we will want to treat them as individual absorbers or combine them in our statistical sample. For example, if the metallicities of paired absorbers are always analagous, then it will show there is not much metallicity variation along the line of sight in redshift space. On the other hand, if some paired absorbers have very different metallicities, it will indicate metallicity variation on a small redshift scale and we should not treat these paired absorbers as single absorbers. 

In Tables~\ref{t-prox} and \ref{t-close}, we list the proximate and paired absorbers, respectively. The immediate inference from these tables is that the effect (if any) of these absorbers on the global properties of the absorbers in our sample must be small since there are only  13 proximate absorbers and 30-paired absorbers in our total sample of \ssz\ absorbers. Out of the 13 proximate absorbers, 8 are also paired absorbers, i.e., proximate absorbers with two closely redshift separated absorbers are more common (see Table~\ref{t-close}). This is likely caused by the larger overdensity around QSOs, which can create an excess of \hi\ absorption; such  enhanced \hi\ \lya\ opacity  has been observed toward $z\sim 2$ QSOs  \citep[e.g.,][]{hennawi06,prochaska13}. Visually, the overall atomic and ionic velocity profiles of the proximate absorbers do not appear obviously different than the intervening absorbers ($\Delta v > 3000$ \km). The only exception is the absorber at $z=0.470800$ toward J161916.54+334238.4 where there is extremely strong absorption of intermediate to high ions; in particular, strong absorption features of \svi, \sv, and \siv\ are not typically observed except in the proximity of a very hard ionizing source, implying that this particular absorbers is likely an associated system. 

Ten of the 30 paired absorbers have 3 (and one has 4) closely separated components. The individual \hi\ column densities for these paired absorbers are below $10^{17}$ \cmm, but this is an observational bias.  Owing to the COS resolution, separating absorbers can be done more easily and robustly only for relatively low \nhi\ absorbers. As shown in Fig.~\ref{f-dv90-hi-z} and discussed in \S\ref{s-res-vel}, absorbers with $\mlnhi \ga 17.2$ have typically larger \dv\ than absorbers with lower \nhi, corresponding to more complex velocity profiles with more than one component. In paper II where we determine the metallicity of the absorbers with $16.2 \le \mlnhi < 19 $, we will consider if there is any effect on the comparison of stronger and weaker \nhi\ absorbers. 

Finally, we note that our survey shows unambiguously  how rare are SLFSs, pLLSs, or LLS with $\mdv\ge 200$-500 \km\ and a number of components $\ga 5$--10 at $z<1$. The only known reported cases are two LLSs at $z<1$ \citep{tripp11,muzahid15}. This implies that only $1.5\%$ (90\% confidence level; 10\% of the LLSs if only LLSs are considered) of the absorbers with $15<\mlnhi < 19$ at $z\la 1$ are very broad. This contrasts remarkably from the finding of the KODIAQ survey at high $z$ where, with the same \hi-selection, more than $50\%$ absorbers have $\mdv\ge 200$-500 \km\ and a number of components $\ga 5$--10 at $z\sim 2$--4 (\citealt{lehner14,lehner17}, and see also \citealt{simcoe02,simcoe04}). This  suggests that a larger fraction of these high-$z$ absorbers may probe large-scale outflows from high-redshift galaxies at an epoch where galaxies were forming stars at much higher rates.

\section{Summary}\label{s-summary}
Using the \hst\ COS G130M and G160M archive, we have built the largest sample to date of \hi-selected absorbers with $15<\mlnhi < 19$ at $z\la 1$ for which we can estimate the metallicities. The sample analyzed in this paper has \ssz\ absorbers, and our total sample consists of \tsz\ absorbers (where the additional data were observed \hst\ STIS, \fuse, and \hst\ COS). This survey is about an order of magnitude larger than the first survey of pLLSs and LLSs at $z<1$ we undertook \citepalias{lehner13}, and it is the first survey targeting and analyzing absorbers with $15<\mlnhi < 16.2$ in the same manner as the pLLSs and LLSs. More quantitatively, we increase the sample of pLLSs by a factor 2 (from 44 in \citetalias{wotta16} to 82) and of LLSs by a factor 2.6 (from 11 in \citetalias{wotta16} to 29), and assemble for the first time a sample of 152 SLFSs. The \hi\ selection ensures that no bias is introduced in the metallicity distribution of these absorbers. We consider only absorbers with $\mlnhi >15$ because the spectra of $z<1$ QSO do not have high enough SNRs to sensitively probe metallicity that is $\la 10\%$ solar at lower column densities. For about half of our sample observed with COS G130M and/or G160M (\gbmgii\ absorbers), we have also high resolution Keck HIRES and VLT UVES observations of the strong NUV transition of \mgii\ and \feii. All the measurements and spectra are made available online. 

In subsequent papers, we will present the photoionization models used to determine the metallicity of these absorbers, the metallicity distributions of these absorbers, and the evolution of the metallicities over 7 orders of magnitude in \nhi\  ($15 < \mlnhi \la 22$). We will also study how the properties of the high ions (in particular \ovi) correlate with those of the low ions (paper IV). Here we have presented the basic measurements (column densities and kinematics) from the absorption of these absorbers to empirically characterize some of the properties of gas with $15<\mlnhi < 19$ at $z\la 1$. Our initial main findings are as follows. 

\begin{enumerate}[wide, labelwidth=!, labelindent=0pt]
\item From the comparison of the absorption profiles and the width of the profiles, we conclude low ions (singly and doubly ionized species) and \hi\ trace the same gas. Thus they can be modeled {\it a priori} with a single ionization phase model. This does not necessarily apply to the higher ions, and  absorbers with high ion absorption must be multiphase. From the profile fitting of the \hi\ transitions, we show that on average the gas temperature is cool with $T < 5 \times 10^4$ K, which is consistent with the gas being photoionized. 

\item  The [\mgii/\hi] ratio spans over 2 orders of magnitude at any \nhi\ over the range $15< \mlnhi < 19$, implying a metal enrichment from a  solar to $\la$1/100 solar metallicity. We find that the absorbers are most likely all metal-enriched at some level, with only $<$3\% of the absorbers (90\% confidence level) showing no metal absorption in the spectra. The bimodal metallicity distribution observed for the pLLSs in \citetalias{lehner13} and \citetalias{wotta16} is apparent from the \mgii\ column densities and equivalent widths, with a lack of data in the range $16.2\la \mlnhi \la 17.2$ around a metallicity corresponding to about 10\% solar. This corresponds to the previously observed dip in the metallicity distribution. 

\item There is no strong dependence of the ionic ratios with \nhi. This implies that the ionization properties of the absorbers must not change dramatically over the range  $15<\mlnhi < 19$ (4 orders of magnitude in \nhi). 
 
\item We estimate  $\langle [$\feii/\mgii$]\rangle = -0.4 \pm 0.3$ comparable to that observed in DLAs at any $z$. This ratio can be affected by $\alpha$-enhancement from Type II supernovae, which implies that Fe dust depletion is small if any. There is also no trend of the [\feii/\mgii] ratio with \nhi, which might be expected if dust depletion was a dominant factor. These findings demonstrate that depletion onto dust in these absorbers is negligible. On average we also show that [C/$\alpha$] is consistent with a solar value, but with a large scatter of about 0.3 dex. 

\end{enumerate}
 
\section*{Acknowledgements}

Support for this research was provided by NASA through grant HST-AR-12854 from the Space Telescope Science Institute, which is operated by the Association of Universities for Research in Astronomy, Incorporated, under NASA contract NAS5-26555. This material is also based upon work supported by the  NASA  Astrophysical Data Analysis Program (ADAP) grants NNX16AF52G under Grant No. AST-1212012. This work was supported by a NASA Keck PI Data Award, administered by the NASA Exoplanet Science Institute. Data presented herein were obtained at the W. M. Keck Observatory from telescope time allocated to the National Aeronautics and Space Administration through the agency's scientific partnership with the California Institute of Technology and the University of California. The Observatory was made possible by the generous financial support of the W. M. Keck Foundation. KLC acknowledges support from NSF grant AST-1615296 and appreciates the observational support of K.~Cannoles and T.~Wells, University of Hawai‘i at Hilo undergraduate students. Some of the data presented in this work were obtained from the  Keck Observatory Database of Ionized Absorbers toward QSOs (KODIAQ), which was funded through NASA ADAP grants NNX10AE84G and NNX16AF52G. This research has made use of the Keck Observatory Archive (KOA), which is operated by the W. M. Keck Observatory and the NASA Exoplanet Science Institute (NExScI), under contract with the National Aeronautics and Space Administration. The authors wish to recognize and acknowledge the very significant cultural role and reverence that the summit of Maunakea has always had within the indigenous Hawaiian community.  We wish to acknowledge and thank all the NASA personnel, and in particular the astronauts, who made the servicing mission 4 on the HST a complete success. Their dedication has revolutionized our understanding of the importance of the CGM in the formation and evolution of galaxies and many other fields.

\begin{figure*}[tbp]
\epsscale{1}
\plotone{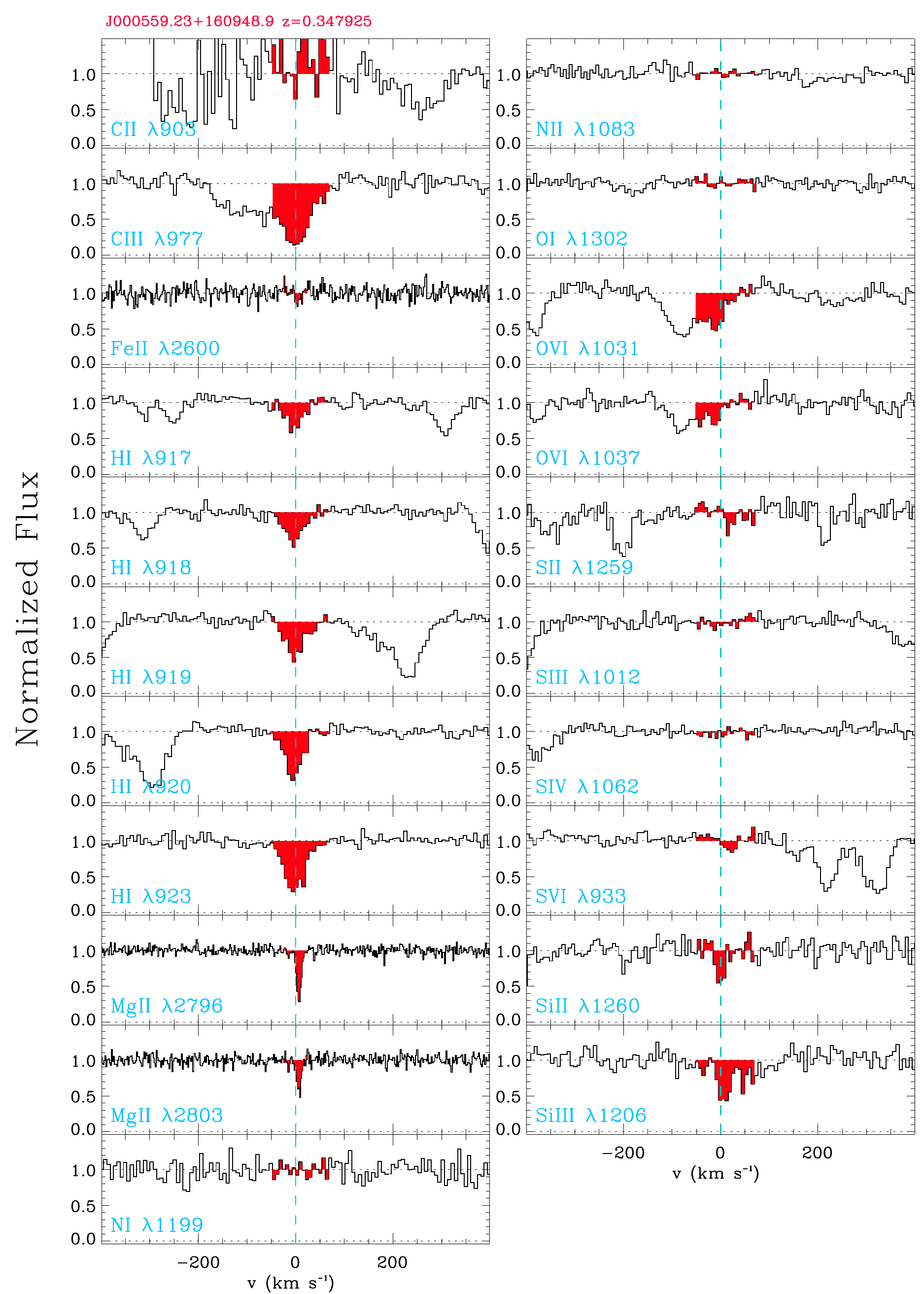}
    \caption{An example of normalized absorption lines as a function of velocity centered on the absorber at $z = 0.347925$ toward J000559.23+160948.9. The EUV and UV transitions are from COS G130M and G160M and \mgiit\  and \feiit\ NUV transitions are in this case Keck HIRES spectra. The red portion in each profile shows the  velocity range of the absorption over which the velocity profile was integrated to derive the column densities, equivalent widths, and kinematics.  The vertical dashed lines mark the zero velocity. Note in this case the simplicity of the velocity profiles in most of the ions and weak \hit\ transitions; only in \ciiit\ and \ovi\ there is evidence of another absorber at $\sim -100$ \km\ (that absorption is only observed in the strongest \hit\ transitions --- \lya\ and \lyb, implying $\mlnhi \la 14$ for that absorber). 
 \label{f-ex-abs-spec}}
\end{figure*}

\appendix
In this Appendix, we provide some information regarding the supplemental files. First, for each absorber studied in this paper, we produced a figure as shown in Fig.~\ref{f-ex-abs-spec} where we plot the normalized profiles of metals and weak \hi\ transitions for which we estimated the column densities. The red portion in each profile shows the  velocity range of the absorption over which the velocity profile was integrated to derive the column densities and kinematics (average velocity and line-width).  The vertical dashed lines mark the zero velocity. 

Second we provide the results in a machine-readable format table. The first two columns provide the J name and the \hst\ name. Columns 3 and 4 give the redshift and its error; column 5 gives the ion; columns 6 and 7 the minimum and maximum velocities for the integration range, columns 8 and 9 the average velocity and its error, columns 10, 11, 12 the column densities and upper and lower errors. The last 3 columns in that file correspond to detection, upper and lower limits (flag\,$ =0,-1,-2$, respectively), reliability of the detection depending if a single or several transitions were detected (1: results based on several transitions or a non-detection, which is always reliable since it would be estimated in an uncontaminated region of the spectrum; 2: results based on at least two transitions, but where only one transition is detected at the $2\sigma$ level and the upper limits agree with that detection; 3: results based only on a single transition or several saturated transitions), and an unique identification number of the absorber.  In that table, a ``999.0'' value in the average velocity and error columns corresponds to no absorption observed at the $2\sigma$ level, and hence that entry is an upper limit on the column density.

\clearpage 
\startlongtable
\begin{deluxetable*}{llccccccl}
\tablecaption{Sample of Targets \label{t-general}}
\tablehead{
\colhead{Name} & \colhead{\hst\ Name} & \colhead{$z_{\rm em}$} & \colhead{$\lambda_{\rm min}$} & \colhead{$\lambda_{\rm max}$} & \colhead{SNR} & \colhead{PID} & \colhead{$N_{\rm abs}$}& \colhead{Obs.}\\
\colhead{} & \colhead{} & \colhead{} & \colhead{(\AA)} & \colhead{(\AA)}& \colhead{}& \colhead{}& \colhead{}& \colhead{} 
}
\startdata
\hline
   J000024.42-124547.7 & PHL2525                   &  0.20000 &   1132.6 &   1760.0 &   18.0 & 12604               &  0 & \nodata  \\
   J000559.23+160948.9 & PG0003+158                &  0.45045 &   1132.7 &   1792.2 &   24.2 & 12038               &  3 & UVES     \\
   J001224.02-102226.3 & SDSSJ001224.01-102226.5   &  0.22822 &   1135.4 &   1795.5 &    5.4 & 12248               &  0 & \nodata  \\
   J002330.58+154744.9 & J002330.58+154744.9       &  0.41041 &   1144.0 &   1460.5 &    4.6 & 14071               &  0 & \nodata  \\
   J004222.29-103743.8 & SDSSJ004222.29-103743.8   &  0.42397 &   1135.5 &   1795.5 &    6.5 & 11598               &  1 & HIRES    \\
   J004705.89+031954.9 & PG0044+030                &  0.62400 &   1171.0 &   1467.0 &    7.6 & 12275               &  3 & UVES     \\
   J010131.15+422935.4 & 2MASX-J01013113+4229356   &  0.19000 &   1135.4 &   1748.3 &    7.4 & 11632               &  0 & \nodata  \\
   J011013.14-021952.8 & LBQS-0107-0235            &  0.96000 &   1132.8 &   1800.0 &   10.8 & 11585               &  4 & UVES     \\
\enddata
\tablecomments{Each of these targets was observed with COS G130M and/or G160M. SNR is given in for COS spectra binned by 3 pixels per resolution element. $N_{\rm abs}$ is the number of absorbers along a given sightline where we could estimate reliable column densities and metallicities. If a number is between parentheses in the column $N_{\rm abs}$, it means we could only derive a lower limit on \nhi, except for J131956.23+272808.2 where the absorber at $z=0.660356$ was originally selected based on the presence of \mgiit\ (see text for more details); in all these cases, these absorbers are not included further in our \hit-selected survey.}
\end{deluxetable*}

\clearpage 
\startlongtable
\begin{deluxetable}{lcccc}
\tablecaption{Atomic Parameters and Abundances for the Surveyed atoms and ions \label{t-ion-list}}
\tablehead{
\colhead{Species} & \colhead{$\lambda$} & \colhead{$\log \lambda f$} & \colhead{$A_\odot$}& \colhead{Ion. Flg.}\\
\colhead{} & \colhead{(\AA)} & \colhead{}& \colhead{}& \colhead{}
}
\startdata
\hline
\hit\        &  1215.6700 &     2.70 &  12.00 & 1 \\
\hit\        &  1025.7222 &     1.91 &  12.00 & 1 \\
\hit\        &   972.5367 &     1.45 &  12.00 & 1 \\
\hit\        &   949.7430 &     1.12 &  12.00 & 1 \\
\hit\        &   937.8034 &     0.86 &  12.00 & 1 \\
\hit\        &   930.7482 &     0.65 &  12.00 & 1 \\
\hit\        &   926.2256 &     0.47 &  12.00 & 1 \\
\hit\        &   923.1503 &     0.31 &  12.00 & 1 \\
\enddata
\tablecomments{\hit\ and metal ions and atoms that were systematically surveyed. Solar abundances are from \citet{asplund09}. Atomic parameters for the FUV  and EUV lines are from \citet{morton03} and \citet{verner94}. respectively.  The table is sorted in decreasing order of abundances. The last column present the ``ionization flag'', i.e., the elements that were used or not in the photoionization models in papers II and III: 1: these are $\alpha$-elements with low ionization stages that are always used if uncontaminated; 2: always used  if uncontaminated but allowing for possible variation in the [C\,$/\alpha$] ratio; 3: often used, but sometimes removed because nucleosythesis or depletion can affect the  N\,$/\alpha$ or  Fe\,$/\alpha$ ratio, respectively; 4: sometimes used dependending on the ionization conditions as these ions can be produced by multiple ionization processes; 5: never included because their ionization stages are too high to typically be produced by the same processes or at the same densities than observed for lower ions and \hit. }
\end{deluxetable}

\clearpage 
\startlongtable
\begin{deluxetable}{lccc}
\tablecaption{Average velocities and column densities of the metal ions in the weak \hit\ absorbers \label{t-results}}
\tablehead{
\colhead{Species} & \colhead{$[v_1,v_2]$} & \colhead{$\langle v \rangle$} & \colhead{$\log N$}\\
\colhead{} & \colhead{(\km)} & \colhead{(\km)}& \colhead{[cm$^{-2} $]}
}
\startdata
\hline
\multicolumn{4}{c}{J000559.23+160948.9, $z= 0.305789$ } \\
\hline
\ciit\ $\lambda$903       &  $ -90, 70 $    &  \nodata             &                   $ < 12.83 $ \\
\ciiit\ $\lambda$977      &  $ -90, 70 $    &  $ 34.5 \pm 8.8 $    &             $12.93 \pm 0.07 $ \\
\feiit\ $\lambda$2600     &  $ -25, 25 $    &  \nodata             &                   $ < 12.40 $ \\
\feiiit\ $\lambda$1122    &  $ -90, 70 $    &  \nodata             &                   $ < 13.43 $ \\
\hit\ $\lambda$926        &  $ -90, 70 $    &  $ -5.7 \pm 8.2 $    &             $15.35 \pm 0.09 $ \\
\hit\ $\lambda$937        &  $ -90, 70 $    &  $ -7.3 \pm 3.1 $    &             $15.30 \pm 0.03 $ \\
\hit\ $\lambda$949        &  $ -90, 70 $    &  $ -7.0 \pm 1.5 $    &             $15.35 \pm 0.02 $ \\
\enddata
\tablecomments{Upper limits ($<$) are non-detections quoted at the 2$\sigma$ level. Column densities preceded by $>$ are lower limits owing to saturation in the absorption (lower limit values were increased by 0.15 dex to additionally correct for saturation, see text for more detail). For a given atom or ion with more than one transition, we list in the row with no wavelength information the adopted weighted average column densities and velocities. }
\end{deluxetable}
\clearpage 
\startlongtable
\begin{deluxetable}{lccc}
\tablecaption{Average velocities and column densities of the metal ions from the L13 sample \label{t-results-l13}}
\tablehead{
\colhead{Species} & \colhead{$[v_1,v_2]$} & \colhead{$\langle v \rangle$} & \colhead{$\log N$}\\
\colhead{} & \colhead{(\km)} & \colhead{(\km)}& \colhead{[cm$^{-2} $]}
}
\startdata
\hline
\multicolumn{4}{c}{J015513.20-450611.9, $z= 0.225958$ } \\
\hline
\ciit\ $\lambda$1036      &  $ -100, 100 $  &  $ -2.6 \pm 2.9 $    &             $14.08 \pm 0.03 $ \\
\ciit\ $\lambda$1334      &  $ -100, 100 $  &  $ 14.4 \pm 3.9 $    &             $14.05 \pm 0.04 $ \\
\ciit\                    &                 &  $ 3.6 \pm 2.3 $     &             $14.07 \pm 0.02 $ \\
\ciiit\ $\lambda$977      &  $ -100, 100 $  &  $ 2.9 \pm 0.9 $     &                   $ > 14.00 $ \\
\feiit\ $\lambda$2600     &  $ -50, 70 $    &  \nodata             &                   $ < 12.25 $ \\
\hit\ $\lambda$930        &  $ -100, 100 $  &  $ -8.9 \pm 3.2 $    &                   $ > 16.24 $ \\
\hit\ COG                 &  \nodata        &  \nodata             &             $16.79 \pm 0.08 $ \\
\enddata
\tablecomments{Upper limits ($<$) are non-detections quoted at the 2$\sigma$ level. Column densities preceded by $>$ are lower limits owing to saturation in the absorption (lower limit values were increased by 0.15 dex to additionally correct for saturation, see text for more detail). For a given atom or ion with more than one transition, we list in the row with no wavelength information the adopted weighted average column densities and velocities. }
\end{deluxetable}

\clearpage 
\begin{deluxetable}{lcc}
\tablecaption{Sample of Absorbers not Included in our Sample \label{t-hifail}}
\tablehead{
\colhead{Name} & \colhead{$z_{\rm abs}$} & \colhead{\lnhi} \\
\colhead{} & \colhead{} & \colhead{([\cmm])} 
}
\startdata
\hline
J100102.64+594414.2  &   0.303648  &   $ >  18.00	      $    \\  
J102117.47+343721.6  &   0.621005  &   $ >  17.65	      $    \\  
J102416.73+242211.6  &   0.522433  &   $ >  18.10	      $    \\  
J113327.78+032719.1  &   0.237384  &   $ >  17.80	      $    \\  
J123647.72+060048.4  &   0.540983  &   $ >  18.05	      $    \\  
J124035.51+094941.0  &   0.494686  &   $ >  18.05	      $    \\  
J124410.88+172104.6  &   0.550600  &   $ >  17.85	      $    \\  
J125224.99+291321.1  &   0.411039  &   $ >  18.00	      $    \\  
J131956.23+272808.2  &   0.660356  &   $   18.59\ \pm 0.05    $    \\  
J134251.60-005345.3  &   0.227331  &   $ >  17.85	      $    \\  
J161649.42+415416.3  &   0.320992  &   $ >  17.90	      $    \\  
J213135.20-120704.5  &   0.429789  &   $ >  18.10	      $    \\  
\enddata
\tablecomments{All the absorbers with ``$>$" in the \hit\ column density column are lower limits, and hence absorbers for which we cannot reliably determine the metallicity.  For J131956.23+272808.2, we derived a new \hit\ column density based on the COS FUV and NUV data, but that absorber was originally \mgii-selected and is therefore not included in our sample. }
\end{deluxetable}

\begin{deluxetable}{lcc}
\tablecaption{Summary of the mean, standard deviation, and median of $\Delta v_{90}$ \label{t-dv90}}
\tablehead{
\colhead{Species} & \colhead{$\langle \Delta v_{90} \rangle $}  & \colhead{$\Delta v_{90}^{1/2}$}\\
\colhead{} & \colhead{(\km)} & \colhead{(\km)} 
}
\startdata
\hline
\hit\     & $ 84 \pm 27 $ &  78 \\
\ciit\    & $ 81 \pm 32 $ &  76 \\
\ciiit\   & $ 79 \pm 27 $ &  78 \\
\mgii\    & $ 36 \pm 22 $ &  30 \\
\siiiit\  & $ 88 \pm 26 $ &  82 \\
\enddata
\end{deluxetable}

\begin{deluxetable}{lccc}
\tablecaption{Proximate Absorbers \label{t-prox}}
\tablehead{
\colhead{Name} & \colhead{$z_{\rm em}$}& \colhead{$z_{\rm abs}$} & \colhead{$\Delta v$}\\
\colhead{} & \colhead{} & \colhead{} & \colhead{(\km)} 
}
\startdata
\hline
J011935.69-282131.4    &  0.34886 & 0.348326 & $   118.7 $ \\
J011935.69-282131.4    &  0.34886 & 0.348731 & $    28.7 $ \\
J100110.20+291137.5    &  0.55800 & 0.556468 & $   295.1 $ \\
J110047.85+104613.2    &  0.42199 & 0.415099 & $  1459.9 $ \\
J122454.44+212246.3    &  0.43351 & 0.420788 & $  2684.4 $ \\
J122454.44+212246.3    &  0.43351 & 0.421357 & $  2563.3 $ \\
J124511.26+335610.1    &  0.71700 & 0.712976 & $   704.3 $ \\
J142735.59+263214.6    &  0.36600 & 0.366054 & $   -11.9 $ \\
J155048.29+400144.9    &  0.49684 & 0.491977 & $   977.1 $ \\
J155048.29+400144.9    &  0.49684 & 0.492632 & $   845.2 $ \\
J155232.54+570516.5    &  0.36600 & 0.366020 & $    -4.4 $ \\
J155232.54+570516.5    &  0.36600 & 0.366521 & $  -114.3 $ \\
J161916.54+334238.4    &  0.47093 & 0.470800 & $    26.5 $ \\
\enddata
\tablecomments{Absorbers with $\Delta v \equiv (z_{\rm em} - z_{\rm abs})/(1+z_{\rm abs})\, c < 3000$ \km.}
\end{deluxetable}

\begin{deluxetable*}{lcccl}
\tablecaption{Closely Redshift Separated Absorbers \label{t-close}}
\tablehead{
\colhead{Name} & \colhead{$[\mlnhi^1,\mlnhi^2]$} & \colhead{$[z^1_{\rm abs},z^2_{\rm abs}]$} & \colhead{$\Delta v_c$} & \colhead{Type}\\
\colhead{} & \colhead{[\cmm]} & \colhead{}  & \colhead{(\km)}& \colhead{} 
}
\startdata
\hline
J004705.89+031954.9    & $[15.57 , 15.56]$ & $[0.313870 , 0.314257]$ &     88.3     & Int. \\
J011935.69-282131.4    & $[15.65 , 16.03]$ & $[0.348326 , 0.348731]$ &     90.1     & Prox \\
J020157.16-113233.1    & $[15.63 , 15.47]$ & $[0.322565 , 0.323073]$ &    115.2$^a$ & Int. \\
J020157.16-113233.1    & $[15.47 , 16.09]$ & $[0.323073 , 0.323404]$ &     75.0$^a$ & Int. \\
J020157.16-113233.1    & $[16.09 , 15.84]$ & $[0.323404 , 0.324507]$ &    249.9$^a$ & Int. \\
J035128.56-142908.0    & $[16.38 , 16.55]$ & $[0.356924 , 0.357173]$ &     55.0     & Int. \\
J044011.90-524818.0    & $[16.20 , 15.66]$ & $[0.614962 , 0.615662]$ &    129.9     & Int. \\	  
J084349.47+411741.6    & $[17.00 , 16.17]$ & $[0.532565 , 0.533507]$ &    184.3     & Int. \\
\enddata
\tablecomments{Absorbers with $\Delta v_c \equiv (z^2_{\rm abs} - z^1_{\rm abs})/(1+z^1_{\rm abs})\, c < 500$ \km\ along the same sightline. $^{a,b,c,d}$: sightlines with 3--4 closely redshift separated absorbers. The last column indicates if the absorber is an intervening or proximate system.}
\end{deluxetable*}

\end{document}